\documentclass[aps, prd, showkeys,superscriptaddress, nofootinbib, floatfix]{revtex4-2}

\usepackage{amsmath}
\usepackage{graphicx}
\usepackage{dcolumn}
\usepackage{bm}

\usepackage[bookmarksnumbered, pdfpagelabels=true, plainpages=false, colorlinks=true, linkcolor=blue, citecolor=blue, urlcolor=blue]{hyperref}

\usepackage{amssymb}
\usepackage{dsfont}
\usepackage{url}
\usepackage{caption,subcaption}
\usepackage{physics}
\usepackage{xcolor}

\begin{document}

\title{Branch-cut in the shear-stress response function of massless $\lambda \varphi^4$ with Boltzmann statistics}

\author{Gabriel S. Rocha}
\email{gabriel.soares.rocha@vanderbilt.edu}
\affiliation{Department of Physics and Astronomy, Vanderbilt University, 6301 Stevenson Science Center
Nashville, TN 37212, USA}
\author{Isabella Danhoni}
\email{idanhoni@theorie.ikp.physik.tu-darmstadt.de}
\affiliation{Institut für Kernphysik,
T. U. Darmstadt, Schlossgartenstraße 2,
D-64289 Darmstadt, Germany}
\author{Kevin Ingles}
\email{kingles@illinois.edu}
\affiliation{
Illinois Center for Advanced Studies of the Universe \& Department of Physics,\\
University of Illinois Urbana-Champaign, Urbana, IL 61801, USA
}

\author{Gabriel S. Denicol}
\email{gsdenicol@id.uff.br}
\affiliation{Instituto de F\'{\i}sica, Universidade Federal Fluminense, Niter\'{o}i, Rio de Janeiro, 24210-346, Brazil}

\author{Jorge Noronha}
\email{jn0508@illinois.edu}
\affiliation{
Illinois Center for Advanced Studies of the Universe \& Department of Physics,\\
University of Illinois Urbana-Champaign, Urbana, IL 61801, USA
}

\begin{abstract}
Using an analytical result for the eigensystem of the linearized collision term for a classical system of massless scalar particles with 
quartic self-interactions, we show that the shear-stress linear response function possesses a branch-cut singularity that covers the whole positive imaginary semi-axis. This is demonstrated in two ways: (1) by truncating the exact, infinite system of linear equations for the rank-two tensor modes, which reveals the cut touching the origin; and (2) by employing the Trotterization techniques to invert the linear response problem. The former shows that the first pole tends towards the origin and the average separation between consecutive poles tends towards zero as power laws in the dimension of the basis. The latter allows one to obtain the response function in closed form in terms of Tricomi hypergeometrical functions, which possess a branch-cut on the above-mentioned semi-axis. This suggests that the presence of a cut along the imaginary frequency axis of the shear stress correlator, inferred from previous numerical analyses of weakly coupled scalar $\lambda \varphi^4$ theories, does not arise due to quantum statistics but instead emerges from the fundamental properties of this system's interactions.
\end{abstract}


\maketitle

\section{Introduction}

The investigation of ultrarelativistic heavy-ion collisions motivates the study of various transport phenomena \cite{Noronha:2024dtq,Florkowski:2017olj,yagi2005quark,heinz2013collective,Arslandok:2023utm,Almaalol:2022xwv,DerradideSouza:2015kpt,Du:2024wjm}. In particular, the description of the quark-gluon plasma phase (QGP), a fluid in which quarks and gluons are deconfined, is performed in terms of relativistic dissipative hydrodynamic models \cite{Denicol:2012cn,Denicol:2014vaa,Schenke:2010nt,Ryu:2015vwa,Paquet:2015lta,Ryu:2017qzn}. Hydrodynamics describes the long-time, long-wavelength behavior of the conserved quantities of a given system. Information on the dynamics is then encoded in the conserved currents of the system, such as the energy-momentum tensor and particle current(s). If the system is in a local equilibrium state, these currents and their evolution are determined by thermodynamic quantities such as temperature, chemical potential, and the four-velocity of the fluid. In more general configurations, additional, dissipative contributions emerge, and various formulations for their evolution can be given (see Ref.~\cite{Rocha:2023ilf} for a recent review). 

The most widely used viscous hydrodynamic formulation in heavy-ion collisions was first proposed by Israel and Stewart \cite{Israel:1976tn,Israel:1979wp}. This formulation assumes that the dissipative currents obey relaxation equations which implement a delayed response to changes in the system's thermodynamic variables. This is to be contrasted with the Navier-Stokes formulation, which considers constitutive relations
connecting dissipative currents to the first-order spatial gradients of the thermodynamic variables in the local rest frame of the fluid, i.e., instantaneous responses to the changes in the fluid. This immediate response leads to equations of motion that are acausal \cite{pichon:65etude} and unstable \cite{hiscock1983stability,hiscock1987linear} in relativistic scenarios. Only recently, has a causal and stable generalization of Navier-Stokes formulation involving first-order derivatives been developed, the so-called BDNK formulation \cite{Bemfica:2017wps,Kovtun:2019hdm,Bemfica:2019knx,Hoult:2020eho,Bemfica:2020zjp}.

Israel-Stewart-like theories possess various transport coefficients beyond the shear and bulk viscosities. The most important ones among those are the relaxation times, which render a causal and stable evolution possible in the linear regime \cite{hiscock1983stability,Pu_2010,Brito:2021iqr,deBrito:2023vzv,Denicol:2021} (for the nonlinear regime, see \cite{Bemfica:2019cop,Bemfica:2020xym}). Furthermore, in Israel-Stewart-like theories, the dissipative currents couple to each other \cite{hiscock1983stability,Denicol:2012cn}, and these couplings are controlled by additional transport coefficients. Only in the asymptotic limit, where the relaxation times are negligible and derivatives are sufficiently small, do the dissipative fluxes in these theories approach the Navier-Stokes constitutive relations.

There is an ongoing debate on the physical interpretation of the Israel-Stewart-like theories, as was recently pointed out in Refs.~\cite{Moore:2018mma,Wagner:2023jgq}. In some formulations, these are considered as ultraviolet regulators, whose sole purpose is to provide a causal and stable evolution \cite{Geroch:1995bx,Lindblom:1995gp,Kostadt:2000ty}, at least in the linear regime. Another interpretation is that causality or stability do not provide bounds on these transport coefficients; rather, the relaxation times should behave in such a way that the dynamics at small gradients, where Navier-Stokes should be applicable, are accurately described \cite{Bhattacharyya:2007vjd,Baier:2007ix,Florkowski:2017olj} (and \citep{Rocha:2023ilf} Sec.~3.5). In this point of view, interpreting the relaxation time as a second-order correction to the gradient series becomes problematic once nonlinear effects from fluctuations are included because the relaxation times can have non-analytic contributions at lower order in the gradient expansion \cite{Kovtun:2011np}. A third approach is that the Israel-Stewart-like equations of motion and, by extension, their transport coefficients, should embody the underlying microscopic interaction, extending the regime of validity of Navier-Stokes theory to this transient, non-equilibrium regime \cite{Denicol:2011fa,Denicol:2012cn,Denicol:2021,Molnar:2013lta} (and \citep{Rocha:2023ilf} Sec.~5.5). In the context of this third point of view, it has been proposed that the relaxation time of a given dissipative current is associated with a pole of the retarded Green's function governing the underlying dynamics in the linearized regime \cite{Denicol:2011fa}. Namely, the relaxation time is determined by the purely imaginary pole closest to the origin for this function in momentum space. This third point of view is limited by the evidence that the shear-stress retarded Green's function for realistic interactions may possess a branch-cut singularity that touches the origin in frequency space. Indeed, in Ref.~\cite{Moore:2018mma}, which considered a system of bosonic scalar particles with quartic interactions, this was demonstrated to be the case for the two-point function of the shear-stress tensor by numerically integrating an effective Boltzmann description with test functions. In Ref.~\cite{Ochsenfeld:2023wxz}, the same system was investigated using a discretization method which renders the computation of the spectrum of the collision term numerically accessible, thus, giving access to the analytical structure of the retarded Green's function. Furthermore, in Ref.~\cite{Gavassino:2024pgl}, a branch cut was shown to arise even in a relatively simple relaxation-time model for a system in which an ideal fluid is coupled with a non-equilibrated radiation gas.

In the present paper, we show that the shear-stress linear response function has a branch cut along the imaginary axis in the frequency plane that touches the origin in a system of scalar self-interacting particles with quartic interactions, even when Boltzmann statistics are used. This emerges in an \textit{analytical} manner by exploiting the recent result of Ref.~\cite{Denicol:2022bsq}, where the spectrum and eigenfunctions of the linearized collision term were computed in exact form for this system. Therefore, the presence of the branch cut in this system does not stem from Bose statistics effects. Its appearance is due to how the particles interact in this system, with a cross-section that vanishes at high energies. Thus, modes possessing large energies with respect to the temperature, which are not hydrodynamic, can, in principle, be present even at arbitrarily late times. Nevertheless, since it is still possible to recover the hydrodynamic transport coefficients derived in Ref.~\cite{Rocha:2023hts} in the asymptotic low frequency regime, where hydrodynamics should work, we argue that these non-hydrodynamic, long-lived modes remain free-streaming and decouple from the part of the system that ``hydrodynamizes''.

This paper is organized as follows. In Sec.~\ref{sec:linear-resp-th}, we work out the linear response theory for the self-interacting scalar particle system from the Boltzmann equation, considering the shear tensor contribution as the only relevant one. In this Section, we also present results for the eigensystem of the linearized collision term since the basis formed by the eigenfunctions of this operator is of central importance to the following sections. Then, in Sec.~\ref{sec:eigenmode-num}, we analyze numerically the shear response function by truncating the eigenmode matrix to investigate the pole structure with varying truncation order. Furthermore, in Sec.~\ref{sec:analyt-trotter}, we perform an analytical study of the same linear response problem through Trotterization methods, where we show that the shear viscosity and the shear-stress relaxation time as computed in Refs.~\cite{Rocha:2023hts,Denicol:2022bsq} can be recovered by the asymptotic series of the full linear response as $\Omega \to 0$. Concluding remarks are made in Sec.~\ref{sec:Conclusion} and, in Appendix \ref{apn:a-b-coeffs}, we discuss the details of the coefficients relevant to the Trotterization procedure we used.

{\bf Notation:} In the present work, we employ the following Fourier convention
\begin{equation}
\label{eq:fourier-convention}
\begin{aligned}
 &
 f(x,p) = \int \frac{d^{4}q}{(2 \pi)^{4}} e^{i q x}  \widetilde{f}(q,p),
\end{aligned}    
\end{equation}
the mostly minus  $(+,-,-,-)$ metric signature, and natural units so that $\hbar = c = k_{B} = 1$.

\section{Relativistic kinetic theory and hydrodynamics}

In non-equilibrium gases, the Boltzmann equation provides the space-time evolution of the one-particle distribution function, $f(x^{\mu},{\bf p}) \equiv f_{\bf p}$, which can be interpreted as a particle density in phase space. If quantum statistics effects are neglected and only two-to-two processes are considered, it reads \cite{DeGroot:1980dk}
\begin{equation}
\label{eq:Boltzmann}
p^{\mu} \partial_{\mu} f_{\bf p}  
=
\frac{1}{2}\int dK \ dK' \ dP' W_{pp' \leftrightarrow kk'} (f_{{\bf k}}f_{{\bf k}'}  -  f_{{\bf p}}f_{{\bf p}'}), 
\end{equation}
where we introduced the Lorentz-invariant integral measure for on-shell massless particles $dP \equiv d^{3}p/[(2\pi)^3p^{0}] = d^{3}p/[(2\pi)^3\vert {\bf p} \vert]$. Furthermore, $W_{pp' \leftrightarrow kk'} = (2\pi)^{6} s \sigma(s,\Theta) \delta^{(4)}(p+p'-k-k')$, denotes the transition rate, $\sigma(s,\Theta)$ is the differential cross section which depends on the Mandelstam variable $s \equiv (k^{\mu}+k'^{\mu})(k_{\mu}+k'_{\mu}) = (p^{\mu}+p'^{\mu})(p_{\mu}+p'_{\mu})$, which denotes the total center-of-momentum energy squared, and $\Theta$, the angle between the three-momenta ${\bf p}$ and ${\bf k}$ in the center-of-momentum frame. The functional form of the differential cross-section depends on the underlying interaction. In this paper, we shall consider a system composed of massless scalar particles whose dynamics are given by the Lagrangian density,
\begin{equation}
\label{eq:lag-phi4}
\mathcal{L} = \frac{1}{2} \partial_{\mu} \varphi \ \partial^{\mu} \varphi
-
\frac{\lambda \varphi^{4}}{4!}.
\end{equation}
The corresponding total cross section is, at leading order in the coupling constant \cite{Peskin:1995ev}, 
\begin{equation}
\label{eq:cross-sec-phi4}
\begin{aligned}
& \sigma_{T}(s) = \frac{1}{2} \int d\Phi d\Theta \, \sin\Theta \, \sigma(s, \Theta) = \frac{\lambda^2}{32 \pi s} \equiv \frac{g}{s},
\end{aligned}    
\end{equation}
where $\Phi$ is the azimuthal angle in the zero-momentum frame and $g \equiv \lambda^{2}/(32 \pi)$.

For any functional form of the cross-section, the non-equilibrium dynamics provided by the Boltzmann equation is compatible with local conservation laws. In a kinetic theory setting, the particle four-current and the energy-momentum tensor can be identified as 
\begin{equation}
\begin{aligned}
& 
N^{\mu} = \int dP p^{\mu} f_{\bf p},\\
&
T^{\mu \nu} = \int dP p^{\mu} p^{\nu} f_{\bf p}.
\end{aligned}    
\end{equation}
And, indeed,using the properties of the collision term \cite{Denicol:2021}, one can derive
\begin{equation}
\begin{aligned}
&
\partial_{\mu}N^{\mu} = 0, \\
&
\partial_{\mu}T^{\mu \nu} = 0.
\end{aligned}    
\end{equation}
We note that beyond leading (tree level) order in the coupling $\lambda$, particle number changing processes appear in the effective kinetic description and particle number is no longer conserved.

The degrees of freedom of the particle four-current and the energy-momentum tensor can be cast in the following manner in terms of a time-like normalized 4-vector $u^{\mu}$ (with $u^{\mu}u_{\mu} = 1$), 
\begin{equation}
\begin{aligned}
\label{eq:decompos-numu-tmunu}
   N^{\mu} &= n u^{\mu} + \nu^{\mu},  \\
    T^{\mu \nu} &= \varepsilon u^{\mu} u^{\nu} - P \Delta^{\mu \nu} + h^{\mu} u^{\nu} + h^{\nu} u^{\mu} + \pi^{\mu \nu},
\end{aligned}
\end{equation}
where $n$ is the total particle density, $\varepsilon$ is the total energy density, $P$ is the total isotropic pressure, $\nu^\mu$ is the particle diffusion 4-current, $h^\mu$ is the energy diffusion 4-current, and $\pi^{\mu\nu}$ is the shear-stress tensor. By construction, the diffusion 4-currents and the shear stress tensor satisfy the following orthogonality relations, $u_\mu \nu^\mu= u_\mu h^\mu = u_\mu \pi^{\mu\nu} = 0$.
We shall consider linear perturbations around a reference local equilibrium state (denoted with a zero-subscript) in what follows. Then, assuming Landau's prescription for the definition of this state \cite{landau:59fluid}, we have 
\begin{equation}
\begin{aligned}
&
n \equiv n_{0}(\mu, T), \quad \varepsilon \equiv \varepsilon_{0}(\mu,T)
\quad P \equiv P_{0}(\mu, T) + \Pi, 
\end{aligned}    
\end{equation}
which define $\mu$, the chemical potential, and $T$, the temperature, such that the total particle and energy density coincide with that of the local equilibrium. In the above equation, we introduced the thermodynamic pressure, $P_{0}(\mu, T)$, and the bulk viscous pressure, $\Pi$, which corresponds to the non-equilibrium correction to the isotropic pressure. Moreover, in Landau's prescription, the fluid four-velocity is defined such  
\begin{equation}
\begin{aligned}
&    
T^{\mu}_{\ \nu} u^{\nu} \equiv \varepsilon_{0} u^{\mu},
\end{aligned}   
\end{equation}
which implies that an observer, with four velocity $u^{\mu}$ does not detect any energy-diffusion 4-current, $h^{\mu} \equiv 0$. The different components appearing in Eq.~\eqref{eq:decompos-numu-tmunu} can be obtained by projecting $N^{\mu}$ and  $T^{\mu \nu}$ accordingly, along or perpendicularly with respect to $u^{\mu}$. In the particular case of kinetic theory, the components related to the local equilibrium state can be computed by performing integrals in momentum space, thus
\begin{equation}
\begin{aligned}
\label{eq:definitions}
    n_{0} &\equiv u_{\mu}N^{\mu} = \int dP E_{\bf p} f_{0\bf{p}}  , \, \, \varepsilon_{0} \equiv u_{\mu}u_{\nu}T^{\mu\nu} = \int dP E_{\bf p}^{2} f_{0\bf{p}} , \\
    P_{0} &\equiv -\frac{1}{3}\Delta_{\mu\nu}T^{\mu\nu}
    = \frac{1}{3} \int dP \left(- \Delta_{\mu \nu} p^{\mu} p^{\nu}\right) f_{0\bf{p}} , 
\end{aligned}
\end{equation}
where $E_{\bf p} = u_{\mu} p^{\mu}$, $\Delta^{\mu \nu}  \equiv g^{\mu \nu} - u^{\mu} u^{\nu}$, and $f_{0\bf{p}}$ denotes the local equilibrium distribution,
\begin{equation}
\begin{aligned}
&
f_{0\bf{p}} \equiv \exp(\alpha - \beta E_\mathbf{p}),
\end{aligned}    
\end{equation}
where $\beta = 1/T$ is the inverse temperature and $\alpha = \mu/T$ is the thermal potential of the fluid. The remaining dissipative currents read
\begin{equation}
\begin{aligned}
\Pi & \equiv \frac{1}{3} \int dP \left(- \Delta_{\mu \nu} p^{\mu} p^{\nu}\right) (f_{\bf{p}} - f_{0\bf{p}}),
\\
\nu^{\mu} & \equiv \Delta^{\mu}_{\nu} N^{\nu} = \int dP \, p^{\langle \mu \rangle} f_{\textbf{p}}, 
\\
\pi^{\mu\nu} & \equiv \Delta^{\mu\nu}_{\alpha\beta} T^{\alpha\beta}
=
\int dP \, p^{\langle \mu}p^{\nu \rangle} f_{\textbf{p}},
\end{aligned}    
\end{equation}
where we introduced the double symmetric, traceless projection tensor $\Delta^{\mu \nu \alpha \beta}  \equiv \frac{1}{2}\left( \Delta^{\mu \alpha } \Delta^{\nu \beta} + \Delta^{\nu \alpha } \Delta^{\mu \beta} \right) - \frac{1}{3}\Delta^{\mu \nu} \Delta^{\alpha \beta}$ and the notations $A^{\langle \mu \rangle} = \Delta^{\mu}_{\ \nu} A^{\nu}$, $A^{\langle \mu \nu \rangle} = \Delta^{\mu \nu \alpha \beta} A_{\alpha \beta}$, for arbitrary vector and rank-two tensors, $A^{\mu}$ and $A^{\mu \nu}$, respectively. It is also noted that, in general, for massless particles and Landau matching conditions, $\Pi = 0$.

\section{Linear response theory for the shear-stress tensor in kinetic theory}
\label{sec:linear-resp-th}

The non-equilibrium dynamics near local equilibrium is described in kinetic theory by the linearized Boltzmann equation. With the cross-section given by Eq.~\eqref{eq:cross-sec-phi4}, it is expressed as
\begin{equation}
\label{eq:Boltzmann}
p^{\mu} \partial_{\mu} f_{0{\bf p}}
+
p^{\mu} \partial_{\mu} \delta f_{\bf p}  
=
\frac{g}{2}\int dK \ dK' \ dP^{\prime} (2 \pi)^{5} \delta^{(4)}(p+p'-q-q') f_{0{\bf p}} f_{0{\bf p}'} (\phi_{{\bf k}}  + \phi_{{\bf k}'}  -  \phi_{{\bf p}} - \phi_{{\bf p}'})
\equiv
f_{0{\bf p}} \hat{L}\phi_{\mathbf{p}}, 
\end{equation}
where $\phi_{\bf p}  =  \delta f_{\bf p}/f_{0 {\bf p}} = (f_{\bf p} - f_{0{\bf p}})/f_{0 {\bf p}}$ is the relative deviation from local equilibrium. In general, $p^{\mu} \partial_{\mu} f_{0\bf p}$ is expressed in terms of gradients of perturbations in temperature, chemical potential, and fluid four-velocity. In what follows, however, we shall consider that only gradients related to the shear stresses are appreciable, 
\begin{equation}
\label{eq:shear-apprec}
\begin{aligned}
&
p^{\mu} \partial_{\mu} f_{0\bf p}
\simeq
- f_{0\bf p} \beta p^{\langle \mu} p^{\nu \rangle} \sigma_{\mu \nu},
\end{aligned}    
\end{equation}
where $\sigma_{\mu \nu} = \partial_{\langle \mu} \delta u_{\nu \rangle}$ is the shear tensor. Here we denoted $\delta u_{\mu}$ as the 4-velocity perturbations, while we shall refer to the velocity of the unperturbed system as $u_{\mu}$. We note that, in the linear regime, $u_{\mu} \delta u^{\mu} = 0$. Also, all projections are now taken with respect to $u_{\mu}$. 

Another simplifying assumption we make is that non-equilibrium perturbations are \emph{spatially homogeneous}. This  means that space-like gradients of $\delta f_{\bf p}$ in the local rest frame are negligible, and then 
+\begin{equation}
\label{eq:homog-apprec}
\begin{aligned}
&
p^{\mu} \partial_{\mu} \delta f_{\bf p} 
= 
E_{\bf p}u^{\mu}\partial_{\mu}\delta f_{\bf p} 
+
p^{\mu} \nabla_{\mu} \delta f_{\bf p}
\simeq
E_{\bf p}\frac{d}{d\tau}\delta f_{\bf p}, 
\end{aligned}    
\end{equation}
where $d/d\tau \equiv u^{\mu} \partial_{\mu}$, $E_{\bf p} \equiv u_{\mu} p^{\mu}$ and $\nabla_{\mu} = \Delta_{\mu}^{\nu} \partial_{\mu}$ reduces to spatial gradients in the local rest frame of the fluid. 
Thus, given \eqref{eq:shear-apprec} and \eqref{eq:homog-apprec},
\begin{equation}
\label{eq:main-linear-sys_bla}
\begin{aligned}
\frac{d\phi_{ \bf p}}{d\tau} 
-
\frac{1}{E_{\bf p}} \Hat{L} \phi_{ \bf p}
= 
- \frac{\beta}{E_{\bf p}} p^{\langle \mu} p^{\nu \rangle} \sigma_{\mu \nu},
\end{aligned}
\end{equation}
which is an inhomogeneous, linear integro-differential equation for $\phi_{ \bf p}$. This equation describes how the deviation function is driven by inhomogeneities of the velocity field contained in the shear tensor. The spectrum of $(1/E_{\bf p})\Hat{L}$ then determines the time scales over which this relaxation process occurs.

For instance, given a judicious choice of an inner product, the operator $(1/E_{\bf p})\Hat{L}$ is self-adjoint and admits an orthogonal set of eigenfunctions. If we expand $\phi_{\bf p}$ using this basis of orthogonal eigenfunctions, $\Psi_{n, {\bf p}}$, 
$\phi_{\mathbf{p}} = \sum_{n} \Phi_{n} \Psi_{n, {\bf p}}$, where the sum over $n$ is converted to an integral in the case of a continuous spectrum, we then obtain the following solution for the deviation function,
$\phi_{\mathbf{p}} = \sum_{n} \Phi^{(0)}_{n} \Psi_{n, {\bf p}} e^{\lambda_{n} t}$ 
where $\Phi^{(0)}_{n}$ are integration constants and $\lambda_{n}$ is the eigenvalue associated with $\Psi_{n, {\bf p}}$. Hence, the fact that the spectrum is non-positive implies that the modes decay exponentially, with the important exception of the linear subspace generated by the collisional invariants, $1,p^{\mu}$, which are associated with infinite relaxation times. Thus, as we stated, the spectrum of $(1/E_{\bf p})\Hat{L}$ provides the characteristic timescales for which the non-hydrodynamic modes evolve towards equilibrium \cite{Moore:2018mma,reichl:99,cercignani:02relativistic}.

Alternatively, the full evolution of the system towards equilibrium can be accessed through the knowledge of the eigensystem of $\Hat{L}$ (which is in general not the same as that of $E_{\bf p}^{-1}\Hat{L}$) and the interplay of the corresponding eigenfunctions with $E_{\bf p}$, seen as an operator in Hilbert space. We shall pursue the latter approach in Fourier space. In this case, the linearized Boltzmann 
\begin{equation}
\label{eq:main-linear-sys}
f_{0{\bf p}} E_{\bf p} i \Omega \ \widetilde{\phi}_{ \bf p} 
-
f_{0{\bf p}} \Hat{L} \widetilde{\phi}_{ \bf p}
= 
f_{0\bf p} \beta p^{\langle \mu} p^{\nu \rangle} \widetilde{\sigma}_{\mu \nu}
\equiv f_{0\bf p} \widetilde S_{\bf p},
\end{equation}
where $\Omega \equiv u^\mu q_\mu$ and $q_{\mu}$ is the Fourier variable (see Eq.~\eqref{eq:fourier-convention}). 
We note that $\widetilde S_{\bf p}$ is not in the kernel of $(E_{\bf p}i\Omega - \hat L)$, and hence this operator can be inverted to find the following particular solution,
\begin{equation}
\widetilde{\phi}_{ \bf p} = ( E_{\bf p} i \Omega  -  \Hat{L})^{-1}\widetilde{S}_{\bf p}.
\end{equation}
In this case, the time-scales that determine the response of the deviation function due to the shear tensor appear as singularities of the operator $( E_{\bf p} i \Omega  -  \Hat{L})^{-1}$. We shall find that these singularities form a branch cut. Thus, there are other modes, besides the hydrodynamic ones, which are long-lived. The presence of a branch-cut singularity prohibits the determination of a single shear-stress relaxation time using the slowest non-hydrodynamic mode \cite{Denicol:2011fa}.

We proceed by making a judicious choice of basis in momentum space. In the present case, we shall employ the eigenfunctions of the linearized collision term for the system described by the Lagrangian in Eq. \eqref{eq:lag-phi4} (at leading order in $\lambda$) determined in Ref. \cite{Denicol:2022bsq}. In general,
\begin{equation}
\label{eq:eigenvalues-lin-col}
\begin{aligned}
&
\hat{L}\left[ L^{(2\ell+1)}_{n{\bf p}} p^{\langle \mu_{1}} \cdots p^{\mu_{\ell} \rangle} \right]
=
\chi_{n}^{(\ell)} 
L^{(2\ell+1)}_{n {\bf p}} p^{\langle \mu_{1}} \cdots p^{\mu_{\ell} \rangle},
 \\
&
\chi_{n \ell} = - \frac{g}{2} \mathcal{M}\left(\frac{n+\ell-1}{n+\ell+1} + \delta_{n0}\delta_{\ell 0}\right),
\end{aligned}    
\end{equation}
where $L^{(2\ell+1)}_{n\mathbf p} = L^{(2\ell+1)}_n(\beta E_{\mathbf p})$ are associated Laguerre polynomials, which form a complete set of orthogonal polynomials, $\mathcal{M} = (n_{0} \beta)/2$ and $p^{\langle \mu_{1}} \cdots p^{\mu_{\ell} \rangle} \equiv \Delta^{\mu_{1} \cdots \mu_{\ell}}_{\nu_{1} \cdots \nu_{\ell}} p^{\nu_{1}} \cdots p^{\nu_{\ell}}$ are irreducible tensors built from products of the four-momentum, which are completely symmetric, orthogonal to $u^\mu$, and traceless with respect to each pair of indices. The $2 \ell$-rank tensor $\Delta^{\mu_{1} \cdots \mu_{\ell}}_{\nu_{1} \cdots \nu_{\ell}}$, is constructed from combinations of the projection operator, $\Delta^{\mu \nu} = g^{\mu \nu} - u^{\mu} u^{\nu}$, designed to make it symmetric with respect to permutations in any of the indices $\mu_{1} \cdots \mu_{\ell}$ and $\nu_{1} \cdots \nu_{\ell}$, separately, and also traceless within each subset of indices \cite{DeGroot:1980dk,Denicol:2021}. The irreducible tensors satisfy the following orthogonality relation, 
\begin{equation}
\label{eq:main-property-irred-tens}
\begin{aligned}
\int dP\,
p^{\langle \mu_{1}} \cdots p^{\mu_{\ell} \rangle}
p_{\langle \nu_{1}} \cdots p_{\nu_{m} \rangle}
H(E_{\mathbf{p}})
& = \frac{\ell!}{(2 \ell + 1)!!} \Delta^{\mu_{1} \cdots \mu_{\ell}}_{\nu_{1} \cdots \nu_{\ell} } \delta_{\ell m} 
\int dP
\left(\Delta^{\mu \nu} p_{\mu} p_{\nu}\right)^{\ell}
H(E_p), 
\end{aligned}
\end{equation}
where $H(E_p)$ is an arbitrary weight function sufficiently regular so that the integral converges. The set of eigenfunctions obeys the following orthogonality relation
\begin{equation}
\label{eq:orth-laguerre}
\begin{aligned}
&
\int dP \left( \Delta^{\mu \nu} p_{\mu} p_{\nu} \right)^{\ell} L_{n{\bf p}}^{(2 \ell + 1)} L_{m{\bf p}}^{(2 \ell + 1)} f_{0 {\bf p}}
=
A^{(\ell)}_{n} \delta_{nm} 
\equiv
(-1)^{\ell}\frac{n_{0}}{2 \beta^{2 \ell-1}}
\frac{(n+2\ell+1)!}{n!}
\delta_{nm},
\end{aligned}    
\end{equation}
where $n_{0}$ is the local particle density. As it will become clear in what follows, there is a non-trivial aspect in solving equation \eqref{eq:main-linear-sys}, even employing the basis of eigenfunctions of the linear operator $\Hat{L}$. Namely, it is the fact that the operator $\Hat{L}$ and $E_{\bf p}$, regarded as operators in momentum-space functions, do not commute. Indeed, the action of $E_{\bf p}$ on the eigenfunctions $\Hat{L}$ is given by basic properties of associated Laguerre polynomials \cite{NIST:DLMF,gradshteyn2014table}. For instance, taking $\ell =2$, we have   
\begin{equation}
\label{eq:lag-prop-E}
\begin{aligned}
\beta E_{\bf p} L_{n, {\bf p}}^{(5)} 
=
-
(n+1) L_{n+1, {\bf p}}^{(5)}
+
2(n+3) L_{n, {\bf p}}^{(5)}
-
(n+5) L_{n-1, {\bf p}}^{(5)},
\end{aligned}    
\end{equation}
which is a linear combination of eigenfunctions of $\Hat{L}$ with different eigenvalues. Then, acting on an eigenfunction with $\Hat{L}$ and $E_{\bf p}$ leads to different results when these operators are interchanged.

\subsection*{Relaxation time approximation}
\label{sec:RTA}

To gain some physical insight, we first consider the approximation in which
all eigenvalues of $\hat{L}$, for $\ell=2$, are assumed to be equal, that
is, $\chi _{n2}\approx \lim_{n\rightarrow \infty }\chi _{n2}=-g\mathcal{M}%
/2\equiv \chi $. This is not an unreasonable assumption, since the smallest
eigenvalue for $\ell=2$ is given by $\chi _{02}=-g\mathcal{M}/6$ -- just a factor $3$
different from the asymptotic value of $-g\mathcal{M}/2$. This approximation
is equivalent to the relaxation time approximation for $\hat{L}$ for this
interaction, which was shown to display an energy dependent relaxation time $%
\tau_{R}=E_{\mathbf{p}}/\chi $ \cite{Denicol:2022bsq}.

The advantage of this approximation scheme is that the operator $\hat{L}$
will then commute with $E_{\mathbf{k}}$, when applied to the source
term $\tilde{S}_{\mathbf{p}}=\beta p^{\left\langle \mu \right. }p^{\left.
\nu \right\rangle }\tilde{\sigma}_{\mu \nu }$. The solution for the Fourier
transformed deviation function can then be computed explicitly,
\begin{equation}
\tilde{\phi}_{\mathbf{k}}=\frac{1}{i\Omega E_{\mathbf{p}}-\hat{L}}\beta
p^{\left\langle \mu \right. }p^{\left. \nu \right\rangle }\tilde{\sigma}%
_{\mu \nu } \approx \frac{1}{i\Omega E_{\mathbf{p}}-\chi }\beta p^{\left\langle \mu
\right. }p^{\left. \nu \right\rangle }\tilde{\sigma}_{\mu \nu }.
\end{equation}
The shear-stress tensor is then determined to be%
\begin{equation}
\tilde{\pi}^{\mu \nu }=\beta \int dPf_{0\mathbf{p}}p^{\left\langle \mu
\right. }p^{\left. \nu \right\rangle }\frac{1}{i\Omega E_{\mathbf{p}}-\chi }%
p^{\left\langle \alpha \right. }p^{\left. \beta \right\rangle }\tilde{\sigma}%
_{\alpha \beta }\equiv 2\eta \left( \Omega \right) \tilde{\sigma}^{\mu \nu },
\end{equation}
where we defined the response function
\begin{equation}
\label{RTA}
\eta \left( \Omega \right) \equiv \frac{\beta }{15}\int_{0}^{\infty }\frac{%
dE_{\mathbf{p}}}{2\pi ^{2}}f_{0\mathbf{p}}\frac{E_{\mathbf{p}}^{5}}{i\Omega
E_{\mathbf{p}}-\chi }.
\end{equation}

The quantity $\eta \left( \Omega \right) $ corresponds to the shear-stress
response function at vanishing frequency. We see that it displays
singularities at $\Omega =i\left\vert \chi \right\vert /E_{\mathbf{p}%
}=i/\tau _{R}$ and, since the energy is being integrated from zero to
infinity, this will lead to a continuous distribution of singularities that
completely span the upper half of the imaginary axis -- including the region
arbitrarily close to the origin. These long-lived modes that appear near the
origin, in particular, can then be understood as emerging due to high energy
excitations, which relax back to equilibrium extremely slowly as $\tau
_{R}\sim E_{\mathbf{p}}$. As argued in Ref.~\cite{Gavassino:2024pgl}, this occurs because the cross-section of the $\varphi^{4}$ interaction (see Eq.~\eqref{eq:cross-sec-phi4}) is proportional to $1/s$, thus vanishing at asymptotically large energies. The modes that emerge due to small energy excitations will not be long-lived (with the exception of the conserved quantities) and can in principle be treated within the usual schemes applied to understand the emergence of a hydrodynamic regime. This will be discussed further in the remainder of this paper, where we resume the discussion of the full linearized collision operator.

\section{Numerical analysis with a truncated eigenmoment matrix system}
\label{sec:eigenmode-num}

In the next sections, we provide two methods 
to obtain the shear-stress response function without approximating the collision operator
: first, we analyze the system of equations formed by the moments of $f_{\bf p}$ constructed with respect to eigenfunctions of the linearized collision term and, second, we will employ the Trotterization method to solve the linear system. In both cases, we provide strong evidence that the analytical structure of the response function contains a branch cut in $ i 0^{+} < \Omega < i \infty$. 

We first multiply Eq.~(\ref{eq:main-linear-sys}) by the tensor rank-2 eigenfunctions of the linearized collision term, $ L_{n}^{(5)} p^{\langle \alpha} p^{ \beta \rangle} $, and integrate it over momentum, leading to the following moment equations,
\begin{equation}
\label{eq:matrix-probl}
\begin{aligned}
i \frac{\Omega}{\beta} \left[- \Phi_{1}^{\alpha \beta} + 6 \Phi_{0}^{\alpha \beta} \right] 
-
\chi_{02} \Phi_{0}^{\alpha \beta} 
&=
\frac{8 n_{0}}{\beta^{2}} \widetilde{\sigma}^{\alpha \beta}, 
\quad
n = 0, 
\\
i \frac{\Omega}{\beta} \left[- (n+1)\Phi_{n+1}^{\alpha \beta} + 2(n+3) \Phi_{n}^{\alpha \beta} - (n+5) \Phi_{n-1}^{\alpha \beta}\right] 
-
\chi_{n,2} \Phi_{n}^{\alpha \beta} 
&=
0,
\quad n = 1, 2, 3, \cdots
\end{aligned}    
\end{equation}
where we defined the eigenmoments 
\begin{equation}
\label{eq:phi-lag-moments}
\begin{aligned}
&
\Phi_{n}^{\mu \nu}
\equiv
\int dP  L^{(5)}_{n {\bf p}} p^{\langle \mu} p^{ \nu \rangle} \delta f_{\bf p}.
\end{aligned}    
\end{equation}
To obtain Eqs.~\eqref{eq:matrix-probl}, we have employed the fact that the linearized collision term is self-adjoint in this Hilbert space, i.e., given two arbitrary (scalar or tensor) functions of momentum $A_{{\bf p}}$ and $B_{{\bf p}}$, $\int dP
f_{0{\bf p}}A_{{\bf p}}
\Hat{L}B_{{\bf p}}
=
\int dP
f_{0{\bf p}}
B_{{\bf p}}
\Hat{L}A_{{\bf p}}
$ \cite{DeGroot:1980dk}. 
Furthermore, we used identity \eqref{eq:lag-prop-E}. 

From Eq.~\eqref{eq:phi-lag-moments}, it is clear that since $L_{0}^{(5)} = 1$,  one obtains $\Phi_{0}^{\alpha \beta} = \pi^{\alpha \beta}$. Thus, the linear response properties of the shear-stress tensor are completely determined once Eq.~\eqref{eq:matrix-probl} is solved for $\Phi_{0}^{\alpha \beta}$. This system of equations can be cast as, 
\begin{equation}
\begin{aligned}
& 
\sum_{m} \mathcal{S}_{n m} \Phi_{m}^{\alpha \beta} = \mathcal{B}_{n}^{\alpha \beta},
\\
&
\mathcal{S}_{n m} =  - i \widehat{\Omega} (n+1)\delta_{n+1,m} 
+ 
[2(n+3) i \widehat{\Omega} -
\widehat{\chi}_{n}^{(2)}] \delta_{n,m} 
- 
i \widehat{\Omega} (n+5) \delta_{n-1,m},
\quad
n, m = 0, 1, 2, \cdots,
\\
&
\mathcal{B}_{n}^{\alpha \beta} = \frac{16}{g\beta^{3}} \widetilde{\sigma}^{\alpha \beta} \delta_{n,0},
\end{aligned}    
\end{equation}
where we have employed the normalized variables 
\begin{equation}
\label{eq:norm-om-chi}
\begin{aligned}
&
\widehat{\Omega} = \frac{\Omega}{g \beta \mathcal{M}}, 
\quad
\widehat{\chi}_{n}^{(2)} = \frac{\chi_{n,2}}{g \mathcal{M}} = -\frac{1}{2}\frac{n+1}{n+3}.
\end{aligned}    
\end{equation}
The shear-stress linear response problem  is then formally solved by 
\begin{equation}
\label{eq:shear-resp-fun}
\begin{aligned}
&
\pi^{\alpha \beta} =\Phi_{0}^{\alpha \beta} = \sum_{j} (\mathcal{S}^{-1})_{0j} \mathcal{B}_{j}^{\alpha \beta}
\equiv
2 \eta(\widehat{\Omega})
\widetilde{\sigma}^{\alpha \beta},\\
&
\eta(\widehat{\Omega})
=
\frac{8}{g\beta^{3}} (\mathcal{S}^{-1})_{00} . 
\end{aligned}    
\end{equation}
We note that $\mathcal{S}$ is an infinite tridiagonal matrix. For a finite-dimensional truncation of this matrix, $\mathcal{S}_{N}$ (where $N$ is the rank of the matrix), a closed form of the elements of the inverse can be derived \cite{huang1997analytical}. Expressing the matrix $\mathcal{S}$ as 
\begin{equation}
\begin{aligned}
&
(\mathcal{S}_{N})_{nm} = b_{n}\delta_{n+1,m} 
+ 
a_{n} \delta_{n,m} 
+
c_{n} \delta_{n-1,m},
\quad
n, m = 0, 1, 2, \cdots, N-1,
\\
&
a_{n} = 2(n+3) i \widehat{\Omega} + \frac{1}{2} \frac{n+1}{n+3}, \quad 
b_{n} = - i \widehat{\Omega} (n+1), \quad
c_{n} = - i \widehat{\Omega} (n+5),
\end{aligned}
\end{equation}
the $00$ element of the inverse can be expressed as
\begin{equation}
\label{eq:trun-resp-fun}
\begin{aligned}
& 
(\mathcal{S}^{-1}_{N})_{00} = \frac{\varphi_{2}(\widehat{\Omega})}{\theta_{N}(\widehat{\Omega})},
\end{aligned}    
\end{equation}
where $\varphi_{2}$ and $\theta_{N}$ are obtained, respectively, through the following recursion relations
\begin{equation}
\begin{aligned}
&
\varphi_{j} = a_{j-1} \varphi_{j+1} - b_{j-1} c_{j} \varphi_{j+2}, \quad j = N-1, N-2, \cdots, 2, 1 \\
&
\theta_{j} = a_{j-1} \theta_{j-1} - b_{j-2} c_{j-1} \theta_{j-2}, \quad j = 2, 3, \cdots, N, 
\\
&
\varphi_{N+1} = 1, \quad \varphi_{N} = a_{N-1}, \quad \theta_{0} = 1, \quad \theta_{1} = a_{0}.
    \end{aligned}
\end{equation}

\begin{figure}[!ht]
\centering
\begin{subfigure}{0.5\textwidth}
  \includegraphics[scale=0.25]{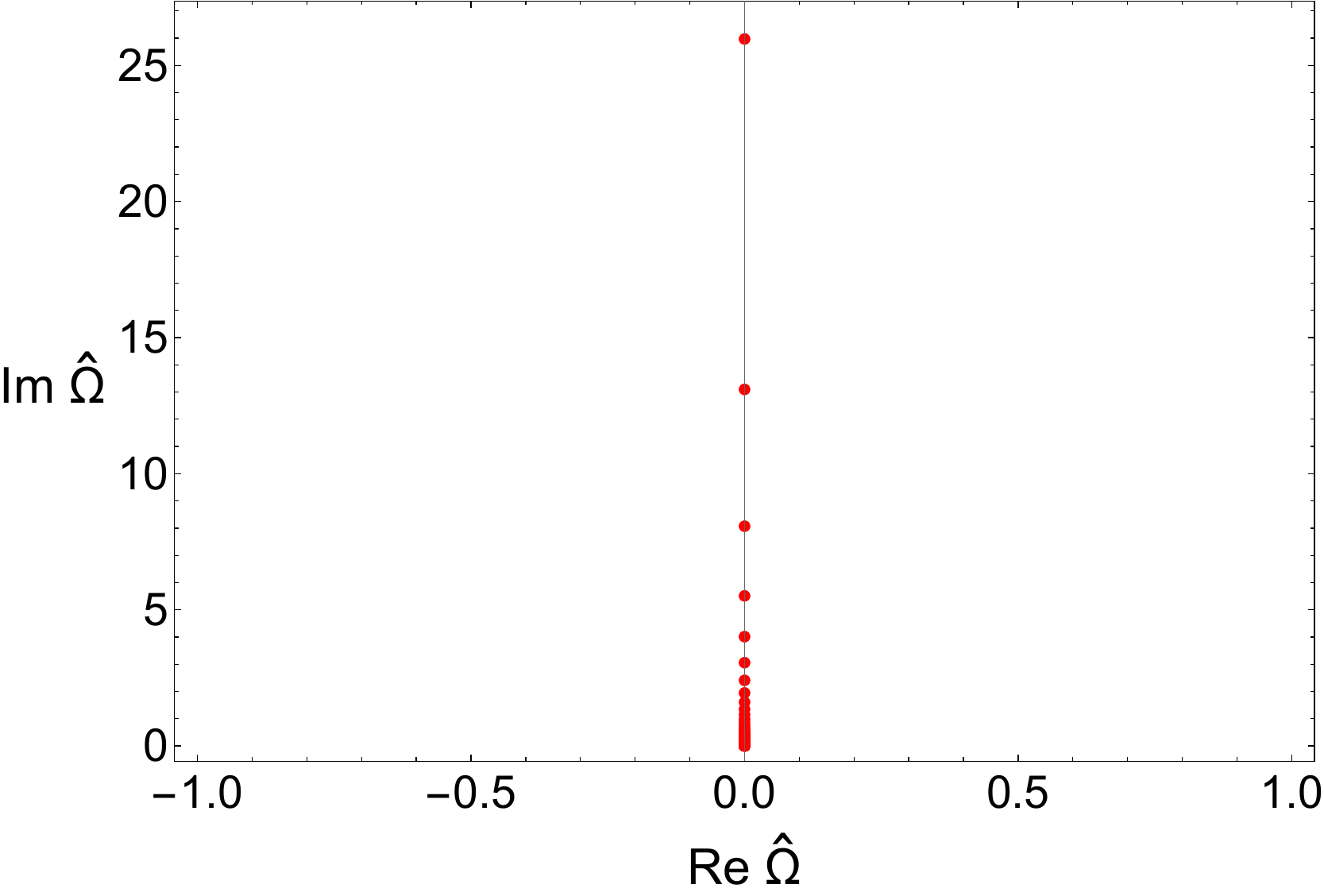} 
\caption{Poles of the response function for $N=1000$}
\label{fig:zeros}
\end{subfigure}\hfil
\begin{subfigure}{0.5\textwidth}
  \includegraphics[scale=0.28]{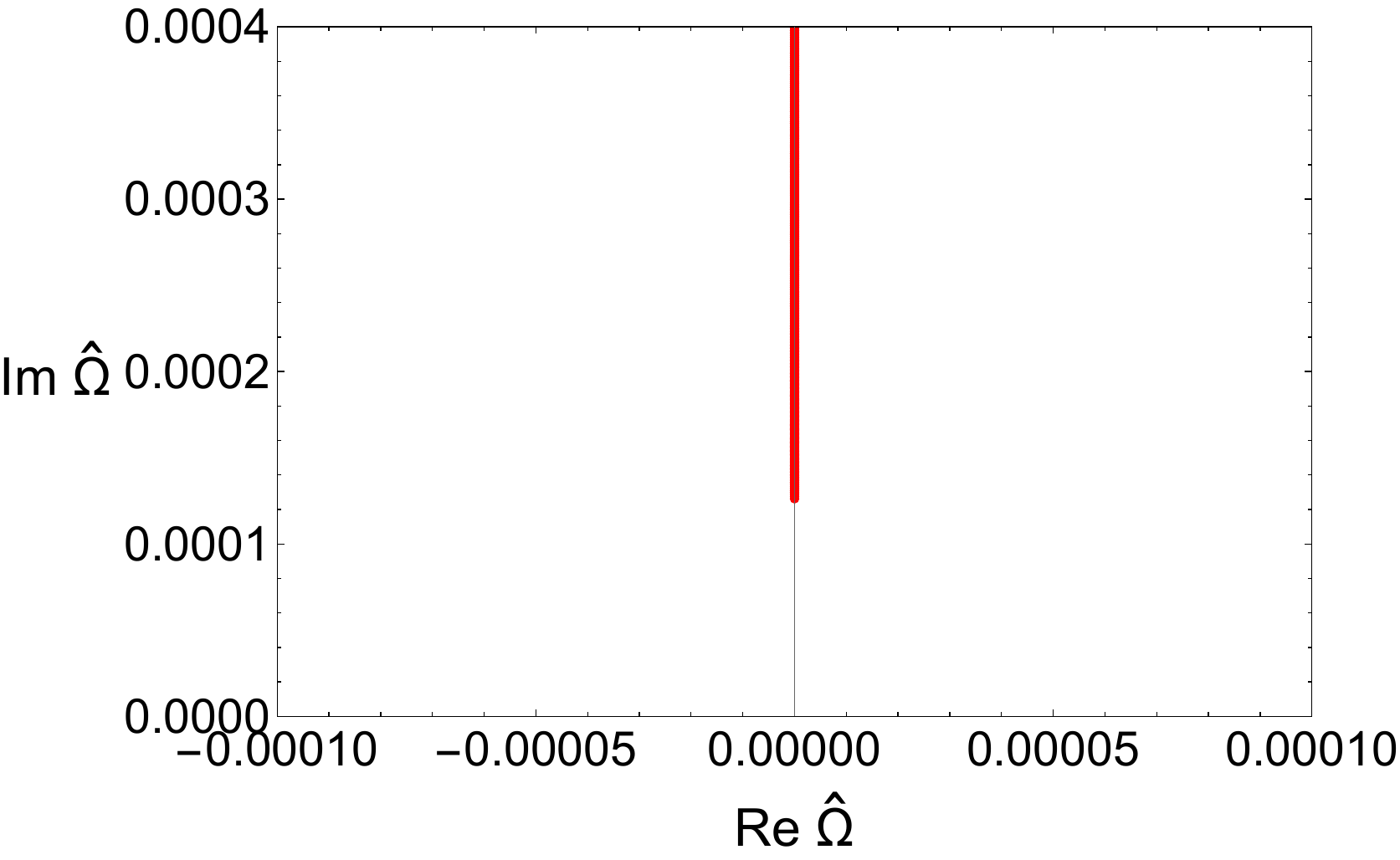}
  \caption{Zoom-in of the plot in panel (a) plot near the origin}
\label{fig:zoom}
\end{subfigure}\hfil
\\
\begin{subfigure}{0.5\textwidth}
  \includegraphics[scale=0.28]{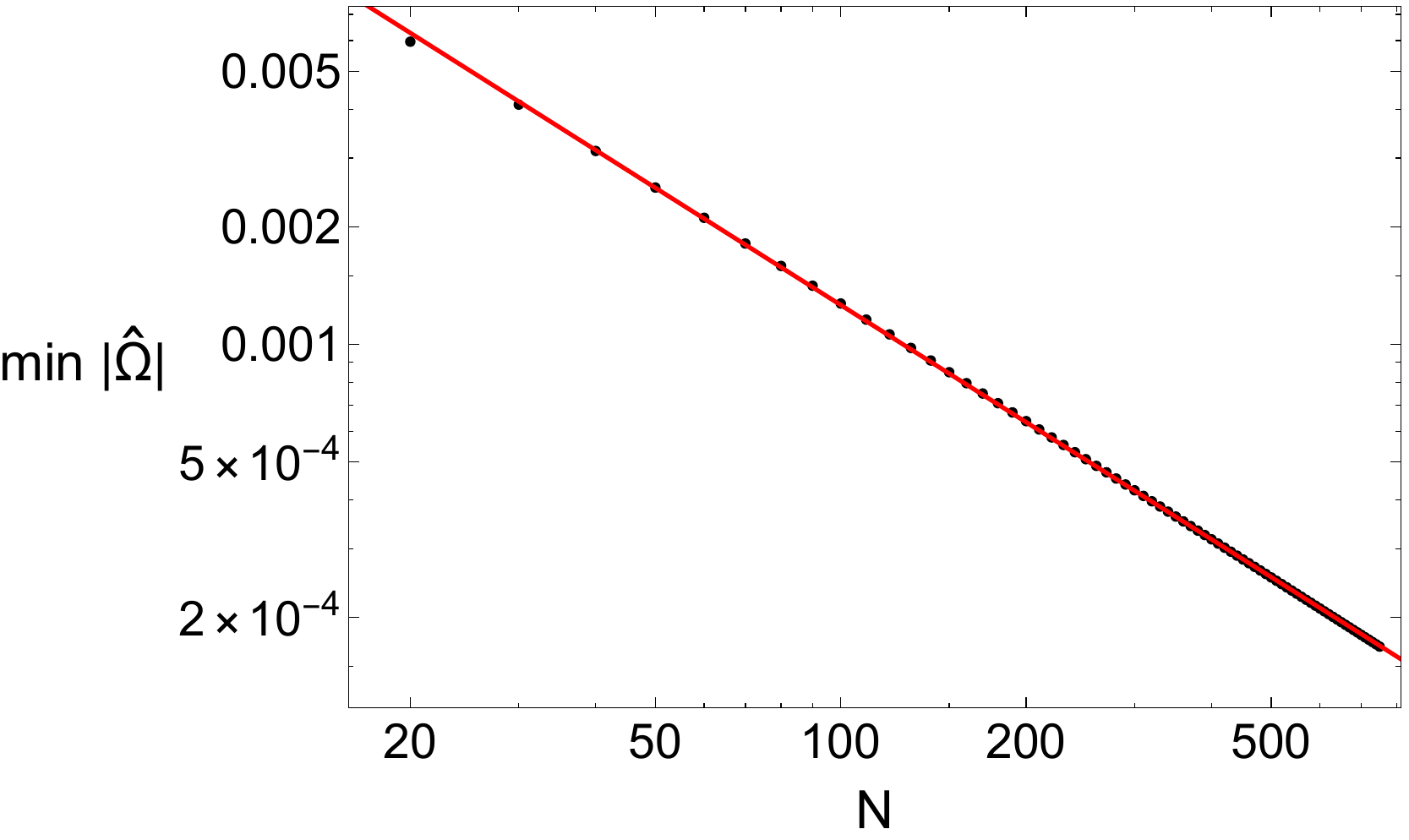} 
\caption{Pole nearest to the origin ($\min|\hat{\Omega}|$) for $N$ up to $N = 750$}
\label{fig:min_poles}
\end{subfigure}\hfil
\begin{subfigure}{0.5\textwidth}
  \includegraphics[scale=0.28]{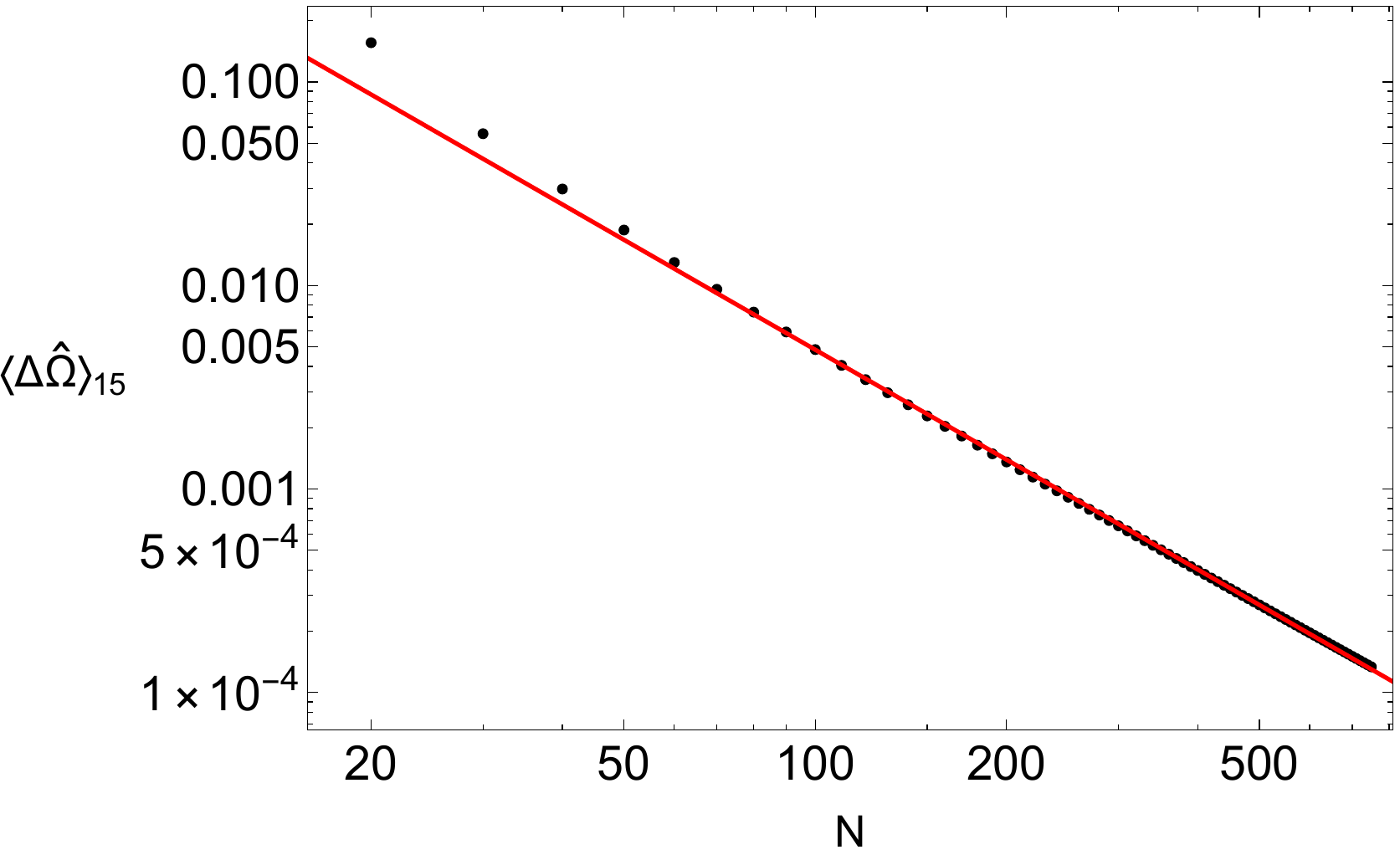} 
\caption{ Average relative distance between the 15 poles closest to the origin $\langle \Delta \Omega \rangle_{15}$}
\label{fig:avg-15-poles}
\end{subfigure}\hfil
\caption{Pole structure information for the truncated response, Eq.~\eqref{eq:trun-resp-fun}, plotted for different matrix rank truncation $N$. }
\label{fig:poles-num}
\end{figure}

Overall, the study of the linear response behavior reduces to the study of the analytical properties of the ratio $\varphi_{2}/\theta_{N}$ as a function of $\widehat\Omega$ in the $N \to \infty$ limit. Indeed, since the roots of $\varphi_{2}(\widehat{\Omega})$ and $\theta_{N}(\widehat{\Omega})$ are different, the poles of the homogeneous Green's function are the zeros of $\theta_{N}$. In Fig.~\ref{fig:poles-num}, we display the main features of the response function obtained from Eq.~\eqref{eq:shear-resp-fun} as the truncation order $N$ becomes large. Figures \ref{fig:zeros} and \ref{fig:zoom} display the poles found for a basis of size $N=1000$. The poles lie on the $\Im \Omega>0$ semi-axis. This is consistent with the fact that these poles stem from the singularities in $\mathcal{S}^{-1}$, which is a proxy to $( E_{\bf p} i \Omega  -  \Hat{L})^{-1}$ that inverts Eq.~\eqref{eq:main-linear-sys}. Since $\Hat{L}$ is a negative semi-definite operator (all its eigenvalues are non-positive), singularities, or zeros in $ E_{\bf p} i \Omega  -  \Hat{L}$, can only arise if\footnote{The singularities emerge in the Im $\widehat{\Omega}>0$ semi-axis due to our conventions of Fourier transform and metric signature.} $\mathrm{Im}{ \Omega} \geq 0$. It is also noted that the poles become closer as $\widehat{\Omega} \to 0$. In Fig.~\ref{fig:zoom}, we display a zoomed-in version of Fig.~\ref{fig:zeros} with the poles closest to the origin. This shows that at this truncation order, $N=1000$, the pole closest to the origin is around $\widehat{\Omega} \simeq 10^{-4}$. In Fig.~\ref{fig:min_poles}, we show the distance of the pole nearest to the origin as $N$ grows up to $N = 750$. The fit to the points shows that the nearest pole approaches the origin as $N$ grows, and $\min \widehat{\Omega} \simeq 1/N^{0.996}$. 

To assess whether the poles converge to a continuous line or a set of discrete points, in Fig.~\ref{fig:avg-15-poles}, we show the average relative distance between the 15 poles closest to the origin, 
\begin{align}
&
\langle \Delta \Omega \rangle_{15} = \sum_{k=1}^{14} \frac{|\omega_{k+1} - \omega_{k}|}{|\omega_{k}|},
\end{align}    
as a function of $N$. Here, $\omega_{k}$ denotes the $k$-th pole closest to the origin (since they all lie in the Re $z$ = 0 axis, $ |\omega_{k+1} - \omega_{k}| = |\omega_{k+1}| - |\omega_{k}|$). The fit shows that $\langle \Delta \Omega \rangle_{15} \sim 1/N^{1.794}$. Thus, the relative distance between the poles closest to $\widehat{\Omega} = 0$ converges to zero faster than the pole closest to the origin converges to $\widehat{\Omega} = 0$. These results indicate the existence of a branch cut in the response function that touches the origin in frequency space as $N\rightarrow \infty$. However, other methods are necessary to assess whether the branch cut extends to the whole Re $\Hat{\Omega}>0$ semi-axis or only to a finite interval. The calculation we performed assuming the relaxation time approximation suggests that this is the case and, indeed, in the next section, we provide further analytical evidence that the cut covers the whole semi-axis.
%
%

\section{Analytical analysis via Trotterization}
\label{sec:analyt-trotter}

As already mentioned, one can formally solve Eq.~\eqref{eq:main-linear-sys} with
\begin{equation}
\label{eq:main-form-sol}
\begin{aligned}
&
\widetilde{\phi}_{ \bf p} = ( E_{\bf p} i \Omega  -  \Hat{L})^{-1}\beta p^{\langle \alpha} p^{\beta \rangle} \widetilde{\sigma}_{\alpha \beta}.
\end{aligned}
\end{equation}
The above equation implies that the linear response of the shear-stress tensor can be expressed as 
\begin{equation}
\label{eq:shear-response-1}
\begin{aligned}
&
\widetilde{\pi}^{\mu \nu}
=
\beta \int dP f_{0 {\bf p}} p^{\langle \mu} p^{\nu \rangle}
 ( E_{\bf p} i \Omega  -  \Hat{L})^{-1} p^{\langle \alpha} p^{\beta \rangle} \widetilde{\sigma}_{\alpha \beta}.
\end{aligned}    
\end{equation}
We note that the difficulty in performing this inversion procedure stems from the fact that $E_{\bf p}$ does not commute with $\Hat{L}$. 
Nevertheless, this inversion can be implemented systematically using Trotterization, a technique which is widely employed in quantum information \cite{wu2002polynomial}. 

Trotterization is based on two identities: first, given an  operator $\Hat{G}$, such that the real part of its eigenvalues are non-negative, its inverse can be expressed as
\begin{equation}
\label{eq:int-1}
\begin{aligned}
&
\Hat{G}^{-1} = \int_{0}^{\infty} dx \ e^{-x G}.
\end{aligned}    
\end{equation}
Second, we employ the Lie-Trotter product formula that expresses the exponential of a sum as the limit of the product of individual exponentials \cite{Hall:2015xtd},
\begin{equation}
\label{eq:int-2}
\begin{aligned}
&
e^{\Hat{A}+\Hat{B}} = \lim_{n \to \infty} (e^{\Hat{A}/n} e^{\Hat{B}/n})^{n}.
\end{aligned}    
\end{equation}
Combining Eqs.~\eqref{eq:int-1} and \eqref{eq:int-2}, identifying $x \Hat{G} = \Hat{A} + \Hat{B}$, $\Hat{A} = - i \Omega x E_{\bf p}$, and $\Hat{B} = x \Hat{L}$ , we can write Eq.~\eqref{eq:shear-response-1} as
\begin{subequations}
\begin{align}
\label{eq:lim-pi-n}
&
\widetilde{\pi}^{\mu \nu} 
=
\lim_{n \to \infty}
\widetilde{\pi}^{\mu \nu}_{n},
\\
&
\label{eq:pi-n}
\widetilde{\pi}^{\mu \nu}_{n} = \beta \widetilde{\sigma}_{\alpha \beta} \int_{0}^{\infty} dx \int dP f_{0 {\bf p}}  \ p^{\langle \mu} p^{\nu \rangle}
\left[ \exp\left(- i \Omega \frac{x}{n} E_{\bf p} \right)  \exp\left( \frac{x}{n} \Hat{L} \right) \right]^{n}
p^{\langle \alpha} p^{\beta \rangle}. 
\end{align}    
\end{subequations}
We note that, since the convergence of the integral of $e^{-x(E_{\bf p} i \Omega  -  \Hat{L})}$ relies on the fact that the spectrum of $\Hat{L}$ is non-positive (since the $E_{\bf p} i \Omega$ term only gives rise to phases) there should be no problem to change the order of the limit and the integral, assuming the former also converges. In the next subsections, we shall make use of the eigenvalues in Eq.~\eqref{eq:eigenvalues-lin-col} and expansions in Laguerre polynomials as a way to provide a closed form of the response function in terms of well-known transcendental functions. From this analytic form, we then infer the analytical properties of the response function. 

\subsection{First Trotterization truncation}

\begin{figure}[!h]
    \centering
\begin{subfigure}{0.5\textwidth}
    \includegraphics[scale=0.8]{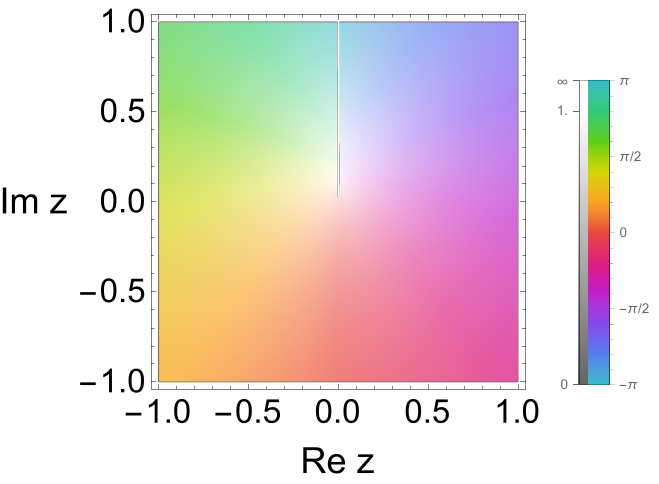}
\end{subfigure}
\caption{Plot in the complex plane of $(iz)^{-6}U(6,6,i/z)$, in terms of which the response function $\eta_{1}( \widehat{\Omega})$ can be expressed.
The colorbar indicates the phase of the Tricomi function, while the gray-scale bar (interpreted as opacity, white being clear and black being opaque) indicates its magnitude.}
\label{fig:U66}
\end{figure}

Taking $n=1$ in Eq.~\eqref{eq:pi-n}, we have that
\begin{equation}
\begin{aligned}
\widetilde{\pi}^{\mu \nu}_{1} &= \beta \widetilde{\sigma}_{\alpha \beta} \int_{0}^{\infty} dx \ \int dP f_{0 {\bf p}}   p^{\langle \mu} p^{\nu \rangle}
\left[ \exp\left(- i \Omega x E_{\bf p} \right)  \exp\left( x \Hat{L} \right) \right]
p^{\langle \alpha} p^{\beta \rangle}
\\
&
=
\frac{2\beta}{15} \widetilde{\sigma}^{\mu \nu}  \int_{0}^{\infty} dx   \int dP \left( \Delta^{\alpha \beta} p_{\alpha} p_{\beta} \right)^{2} f_{0 {\bf p}} \ 
\left[ \exp\left(- i \Omega x E_{\bf p} \right)  \exp\left( x \chi_{02} \right) \right] 
\\
&
=
\frac{8 n_{0}}{\beta^{2}}  \widetilde{\sigma}^{\mu \nu}  \int_{0}^{\infty} dx \frac{\exp\left( x \chi_{02} \right)}{\left(1 + i \Omega \frac{x}{\beta} \right)^{6}}
\equiv
2 \eta_{1}( \widehat{\Omega}) \widetilde{\sigma}^{\mu \nu},
\end{aligned}    
\end{equation}
where, from the first to the second equality, we have used the fact that $\exp\left( x \Hat{L} \right) p^{\langle \alpha} p^{\beta \rangle} = \exp\left( x \chi_{02} \right) p^{\langle \alpha} p^{\beta \rangle}$, which stems from the eigenvalue equation, Eq.~\eqref{eq:eigenvalues-lin-col}, and the orthogonality of the irreducible polynomials \eqref{eq:main-property-irred-tens}. The expression in the second line is very similar to the one obtained with the relaxation time approximation, Eq.~\eqref{RTA} -- this can be seen directly by integrating the second line of the above equation in $x$. The difference is that in the relaxation time approximation, the asymptotic eigenvalue $ \lim_{n\rightarrow \infty }\chi _{n2}$ appears in the expression for the response function, while in the first Trotterization scheme the smallest eigenvalue for $\ell=2$, $\chi _{02}$, appears. We now follow a different path and perform the momentum-space integrals first, going from the second to the third equality in the equation above. We then identify
\begin{equation}
\label{eq:first-trotter-resp-fun}
\begin{aligned}
&
\eta_{1}( \widehat{\Omega}) 
=
\frac{8}{g\beta^{3}} \frac{1}{i \widehat{\Omega}} U\left(1, -4, \frac{|\widehat{\chi}_{02}|}{i \widehat{\Omega}} \right)
=
\frac{8}{g\beta^{3}} 
\frac{|\widehat{\chi}_{02}|^{5}}{(i \widehat{\Omega})^{6}}
 U\left(6, 6, \frac{|\widehat{\chi}_{02}|}{i \widehat{\Omega}} \right)
,
\end{aligned}    
\end{equation}
which is expressed in terms of the dimensionless quantities $\widehat{\Omega}$, $\widehat{\chi}_{n2} $, defined in Eq.~\eqref{eq:norm-om-chi}, and the Tricomi confluent hypergeometric function $U\left(a,b,z\right)$ \cite{NIST:DLMF,bateman_1953,Jr:2021rnt,Erdelyi:1953:HTF1,wolfram:01-tricomi},
\begin{equation}
\begin{aligned}
&
U\left(a,b,z\right)=\frac{1}{\Gamma\left(a\right)}\int_{0}^{\infty} \mathrm{d}t \, e^{-zt}t^{a%
-1}(1+t)^{b-a-1},
\end{aligned}    
\end{equation}
which has the property 
\begin{equation}
\label{eq:U-eq-zn-U}
\begin{aligned}
&
U(a,-n,z) = z^{n+1}U(a+n+1,n+2,z),
\end{aligned}    
\end{equation}
when $n$ is a non-negative integer, which was employed in the last equality of Eq.~\eqref{eq:first-trotter-resp-fun}. The response function $\eta_{1}( \widehat{\Omega})$ can be expressed in terms of the function $(iz)^{-6}U(6,6,i/z)$, whose behavior in complex space is displayed in Fig.~\ref{fig:U66}, where the colors represent the phase of the complex number and their opacity denotes its magnitude. One can see evidence of a singular structure along the $ \Im z > 0$ semi-axis.

Indeed, it is known that the Tricomi confluent hypergeometric function possesses a branch cut in $-\infty < \Re z < 0$, $\Im z =0$, except when $a-b+1 = 0, -1, -2, \cdots$ ($U\left(a,b,z\right)$ reduces to a polynomial in $1/z$) and when $a = 0, -1, -2, \cdots$ ($U\left(a,b,z\right)$ reduces to Laguerre polynomials\footnote{In fact, $L_{n}^{(\alpha)}(x) = [(-1)^{n}/n!] U(-n,\alpha, z)$, for $n = 0,1,2,\cdots$.}) \cite{NIST:DLMF}. In Eq.~\eqref{eq:first-trotter-resp-fun}, we see that $a - b +1 =1$ and $a = 6 > 0$. Thus, since both possibilities mentioned above are false, we conclude that the response function at this order already possesses a branch cut in $0^{+} < \Im \widehat\Omega < \infty$, $\Re \Omega=0$. 

Moreover, it is known that the branch cut discontinuity of the Tricomi function is such that for $z$ in the semi-axis $-\infty < \Re z < 0$, $\Im z =0$ \cite{wolfram:01-tricomi}
\begin{subequations}
\label{eq:bc-disc}
\begin{align}
&
\lim_{\epsilon \to 0+} 
U\left(a,b, z + i \epsilon\right)
=
U\left(a,b, z\right), \\
&
\lim_{\epsilon \to 0+} 
U\left(a,b, z - i \epsilon\right)
=
e^{2 \pi i b}U\left(a,b, z\right)
-
\frac{2 \pi i}{\Gamma(a-b+1) \Gamma(b)} e^{i b \pi} \,_{1} F_{1}(a;b;z),
\end{align}    
\end{subequations}
where $\,_{1} F_{1}(a,b,z) = \sum_{s=0}^{\infty}\frac{{\left(a\right)_{s}}}{{\left(b%
\right)_{s}}s!}z^{s}$  denotes the Kummer confluent hypergeometric function, and  $\left(a\right)_{s} = \Gamma(a+s)/\Gamma(a)$ denotes the Pochhammer symbol.
Hence, we can derive that the response function is also discontinuous along its corresponding cut and, thus,
\begin{equation}
\label{eq:discont-trot-1}
\begin{aligned}
&
\Delta \eta_{1}(\widehat{\Omega}) = \lim_{\epsilon \to 0+} 
\eta_{1}(\widehat{\Omega}) - \eta_{1}(\widehat{\Omega} - \epsilon \widehat{\Omega}^{2})
=
\frac{8}{g\beta^{3}} 
\frac{|\widehat{\chi}_{02}|^{5}}{(i \widehat{\Omega})^{6}}
\left[
U\left(6,6, \frac{|\widehat{\chi}_{02}|}{i \widehat{\Omega}}\right)
-
\frac{i \pi}{60} \,_{1} F_{1}\left(6;6;\frac{|\widehat{\chi}_{02}|}{i \widehat{\Omega}}\right)
\right].
\end{aligned}    
\end{equation}
In what follows, we shall analyze the consequences of the above-mentioned analytical properties on the hydrodynamic limit of the response function.

\subsubsection*{Hydrodynamic series}

The hydrodynamic regime can be defined as the limit in which the gradients are small, or equivalently, the limit where $q^{\mu} \to 0$, or, more specifically for homogeneous perturbations, $\widehat{\Omega} \to 0$. For large $z$-argument, the Tricomi and the Kummer hypergeometric functions behave respectively as \cite{NIST:DLMF}
\begin{equation}
\label{eq:U-large-z}
\begin{aligned}
&
U\left(a,b,z\right)\sim z^{-a}\sum_{s=0}^{\infty}\frac{{\left(a\right)_{s}}{%
\left(a-b+1\right)_{s}}}{s!}(-z)^{-s},
\\
&
\,_{1}F_{1}\left(a,b,z\right) 
\sim
\frac{\Gamma\left(b\right)}{\Gamma\left(a\right)} e^{z}z^{a-b}%
\sum_{s=0}^{\infty}\frac{{\left(1-a\right)_{s}}{\left(b-a\right)_{s}}}{s!}z^{-%
s}.
\end{aligned}    
\end{equation}
This implies that the response function behaves asymptotically as
\begin{equation}
\label{eq:trott-eta-1}
\begin{aligned}
&
\eta_{1}(\widehat{\Omega})
\sim
\eta_{1}(0)
\left[1 + \sum_{s=1}^{\infty} \left(6\right)_{s}\left(-\frac{i \widehat{\Omega}}{|\widehat{\chi}_{02}|}
\right)^{s}
\right]
 ,
\end{aligned}    
\end{equation}
where $\eta_{1}(0) = 48/(g\beta^{3})$ agrees with the expression for shear viscosity computed in Refs.~\cite{Denicol:2022bsq,Rocha:2023hts}. Truncating the above series at next-to-leading order, we have that $\widetilde{\pi}^{\mu \nu} \simeq 2 \eta_{1}(0) [1 - 36 i \widehat{\Omega}] \widetilde\sigma^{\mu \nu}$, which implies that in this regime $[1 + 36 i \widehat{\Omega}]\widetilde{\pi}^{\mu \nu} \simeq 2 \eta_{1}(0) \widetilde\sigma^{\mu \nu}$, which leads to a relaxation equation in real space with $\tau_{\pi} = 72/(g n_{0} \beta^{2})$, that also coincides with the expression computed in Ref.~\cite{Rocha:2023hts} using the method of moments \cite{Denicol:2012cn} and the order-of-magnitude truncation scheme \cite{struchtrup2004stable,Fotakis:2022usk,Wagner:2022ayd}. However, since the coefficients of the series in Eq.~\eqref{eq:trott-eta-1} grow factorially, the radius of convergence of this series is zero. Thus, finite truncations of this gradient series do not improve approximations after an optimal truncation order.

Using the asymptotic expressions \eqref{eq:U-large-z}, the branch cut discontinuity can be re-expressed as,
\begin{equation}
\begin{aligned}
&
\Delta \eta_{1}(\widehat{\Omega})
\sim 
\frac{8}{g\beta^{3}} 
\frac{|\widehat{\chi}_{02}|^{5}}{(i \widehat{\Omega})^{6}}
\left[
-
\frac{i \pi}{60}
\exp{-i\frac{|\widehat{\chi}_{02}|}{\widehat{\Omega}}}
+
\sum_{s=1}^{\infty} \left(6\right)_{s}\left(-\frac{i \widehat{\Omega}}{|\widehat{\chi}_{02}|}
\right)^{s+6}
\right],
\end{aligned}    
\end{equation}
where we notice that since $a = b = 6$ in Eq.~\eqref{eq:discont-trot-1}, only the $s=0$ term in the $\,_{1} F_{1}(a,b,z)$ asymptotic series contributes since $(0)_{s} = \delta_{s,0}$. Thus, one can see that the branch discontinuity has an essential singularity near $\widehat{\Omega} = 0$, manifesting that $\widehat{\Omega} = 0$ is a branch point. The qualitative features of the response function at this Trotterization truncation order shall also be shared by the higher-order ones. The latter, then, shall improve the quantitative agreement with the full response function, which is only accessible as $n \to \infty$.   

\subsection{Second Trotterization truncation}

Inserting $n=2$ in Eq.~\eqref{eq:pi-n}, we have that
\begin{equation}
\begin{aligned}
&
\widetilde{\pi}^{\mu \nu}_{2} = \beta \widetilde{\sigma}_{\alpha \beta} \int_{0}^{\infty} dx \  \int dP f_{0 {\bf p}}  p^{\langle \mu} p^{\nu \rangle}
\left[ \exp\left(- i \Omega \frac{x}{2} E_{\bf p} \right)  \exp\left( \frac{x}{2} \Hat{L} \right) \right]^{2}
p^{\langle \alpha} p^{\beta \rangle},
\end{aligned}    
\end{equation}
where now we need to act the operators on the irreducible moments twice. At this point, the non-commutativity of the $E_{\bf p}$ and $\Hat{L}$ arises more prominently as well as the pattern to compute the $n^\mathrm{th}$ Trotterization truncation. Indeed,
\begin{equation}
\label{eq:second-ord-trot-trun}
\begin{aligned}
\left[ \exp\left(- i \Omega \frac{x}{2} E_{\bf p} \right)  \exp\left( \frac{x}{2} \Hat{L} \right) \right]^{2}
p^{\langle \alpha} p^{\beta \rangle}
&=
\left[ \exp\left(- i \Omega \frac{x}{2} E_{\bf p} \right)  \exp\left( \frac{x}{2} \Hat{L} \right) \right]
\left[ \exp\left(- i \Omega \frac{x}{2} E_{\bf p} \right) p^{\langle \alpha} p^{\beta \rangle}  \right]  \exp\left( \frac{x}{2} \chi_{02} \right)
\\
&
=
\sum_{k_{1}=0}^{\infty} b_{k_{1}}(x, \Omega)\left[ \exp\left(- i \Omega \frac{x}{2} E_{\bf p} \right)  \exp\left( \frac{x}{2} \Hat{L} \right) \right]
 L^{(5)}_{k_{1}, {\bf p}} p^{\langle \alpha} p^{\beta \rangle}  \exp\left( \frac{x}{2} \chi_{02} \right) 
\\
&=
\sum_{k_{1}=0}^{\infty} b_{k_{1}}(x, \Omega)\left[ \exp\left(- i \Omega \frac{x}{2} E_{\bf p} \right) \right]
L^{(5)}_{k_{1}, {\bf p}} p^{\langle \alpha} p^{\beta \rangle}  \exp\left( \frac{x}{2} \chi_{02} + \frac{x}{2} \chi_{k_{1},2} \right) 
\\
&
=
\sum_{k_{1}=0}^{\infty} \sum_{k_{2}=0}^{\infty} b_{k_{1}}(x, \Omega) b_{k_{2}}(x, \Omega)
L^{(5)}_{k_{1}, {\bf p}} L^{(5)}_{k_{2}, {\bf p}} p^{\langle \alpha} p^{\beta \rangle}  \exp\left( \frac{x}{2} \chi_{02} + \frac{x}{2} \chi_{k_{1},2} \right)
\\
&
=
\sum_{k_{1}=0}^{\infty} \sum_{k_{2}=0}^{\infty} \sum_{j = 0}^{k_{1} + k_{2}} a_{j}^{(k_{1}, k_{2})} b_{k_{1}}(x, \Omega) b_{k_{2}}(x, \Omega)
\exp\left( \frac{x}{2} \chi_{02} + \frac{x}{2} \chi_{k_{1},2} \right) L^{(5)}_{j, {\bf p}} p^{\langle \alpha} p^{\beta \rangle}.  
\end{aligned}    
\end{equation}
In the first equality above, we have just employed $\exp\left( x \Hat{L} \right)  p^{\langle \alpha} p^{\beta \rangle} = \exp\left( x \chi_{02} \right) p^{\langle \alpha} p^{\beta \rangle}$, as in the previous subsection. Then, in the second equality, we have used the fact that the Laguerre polynomials form a complete set of functions to express the phases $\exp\left(- i \Omega \frac{x}{2} E_{\bf p} \right)$ as 
\begin{equation}
\label{eq:eiE-b}
\begin{aligned}
&
 \exp\left(- i \Omega \frac{x}{2} E_{\bf p} \right)  = \sum_{k=0}^{\infty} b_{k}(x, \Omega) L_{k, {\bf p}}^{(5)},
\\
&
b_{k}(x, \Omega) = b_{k}\left(\frac{\Omega x}{2 \beta}\right) = \frac{ (i \frac{\Omega x}{2 \beta})^{k} }{\left(1 + i \frac{\Omega x}{2 \beta} \right)^{6+k}}.
\end{aligned}    
\end{equation}
The advantage of this step is seen in the third equality of Eq.~\eqref{eq:second-ord-trot-trun}, where we use $$\exp\left( x \Hat{L} \right) L_{k, {\bf p}}^{(5)} p^{\langle \alpha} p^{\beta \rangle} = \exp\left( x \chi_{k,2} \right) L_{k, {\bf p}}^{(5)}p^{\langle \alpha} p^{\beta \rangle},$$ which also stems from the eigenvalue equation. Then, in the fourth equality of Eq.~\eqref{eq:second-ord-trot-trun}, we express the remaining phase, $\exp\left(- i \Omega \frac{x}{2} E_{\bf p} \right)$, as a linear combination of Laguerre polynomials. This is convenient because expressing the final result in terms of a linear combination of Laguerre polynomials enables us to use the orthogonality of the Laguerre polynomials to perform the momentum space integrals (since $1 = L_{0, {\bf p}}^{(5)}$). This is achieved in the last equality of Eq.~\eqref{eq:second-ord-trot-trun}, where we express the product of Laguerres polynomials as the linear combination of Laguerre polynomials,
\begin{equation}
\label{eq:LL-a-coeffs}
\begin{aligned}
 &
 L_{k_{1}, {\bf p}}^{(5)} L_{k_{2}, {\bf p}}^{(5)}
 =
\sum_{j_{1}=0}^{k_{1} + k_{2}} a^{(k_{1},k_{2})}_{j_{1}} L_{j_{1}, {\bf p}}^{(5)},
\\
&
a^{(k_{1},k_{2})}_{j}
=
(k_{1}+5)!(k_{2}+5)!\sum_{s=0}^{k_{1}}
\sum_{t=j}^{k_{1} + k_{2}} \frac{(-1)^{t+j}}{(k_{1} + k_{2} - t)!(t+10)!}
\left(
\begin{array}{c}
 k_{1} + k_{2} - t \\
 k_{1} - s   
\end{array}
\right)
\left(
\begin{array}{c}
 t + 10\\
 s + 5    
\end{array}
\right)
\left(
\begin{array}{c}
 t \\
 s     
\end{array}
\right)
\left(
\begin{array}{c}
 t+5 \\
 t-j     
\end{array}
\right).
\end{aligned}    
\end{equation}
We note that, in particular, the $a_{j}^{(k_{1}, k_{2})} $ coefficients obey the following properties if any of the indices is zero 
\begin{equation}
\label{eq:a-deltas}
\begin{aligned}
&
a_{j}^{(k_{1}, 0)} 
=
\delta_{j, k_{1}},
\quad
a_{j}^{(0, k_{2})} 
=
\delta_{j, k_{2}},
\quad
a_{0}^{(k_{1}, k_{2})} 
=
\binom{k_{1} + 5}{k_{1}} \delta_{k_{1}, k_{2}}.
\end{aligned}    
\end{equation}
The first two expressions above arise immediately from the fact that if $k_{1} = 0$ (or $k_{2} = 0$), one of the Laguerre polynomials becomes 1, and the expansion becomes trivial. The derivation of the second identity above, as well as details for the computation of the coefficients $a_{j}^{(k_{1}, k_{2})}$ and $b_{k}(x, \Omega)$, are discussed in Appendix \ref{apn:a-b-coeffs}. There, we also argue that, in general, $a_{j}^{(k_{1}, k_{2})} $ is a linear combination of $\delta_{k_{1}, k_{2}-j}$, $\delta_{k_{1}, k_{2}-j+1}$, $\cdots$, $\delta_{k_{1}, k_{2}}$, $\cdots$,  $\delta_{k_{1}, k_{2}+j}$ which is expressed in compact form by Eq.~\eqref{eq:LL-a-coeffs}.

\begin{figure}[!h]
    \centering
\begin{subfigure}{0.5\textwidth}
    \includegraphics[scale=0.7]{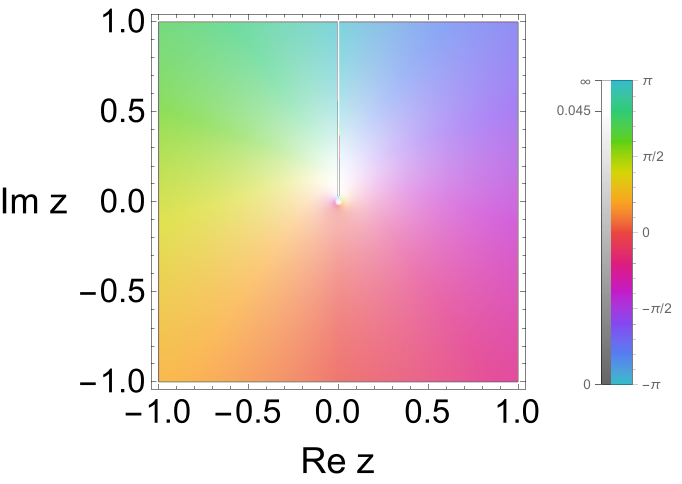}
\caption{$(iz)^{-12}U(13,12,i/z)$}   
\end{subfigure}\hfil
\begin{subfigure}{0.5\textwidth}
    \includegraphics[scale=0.7]{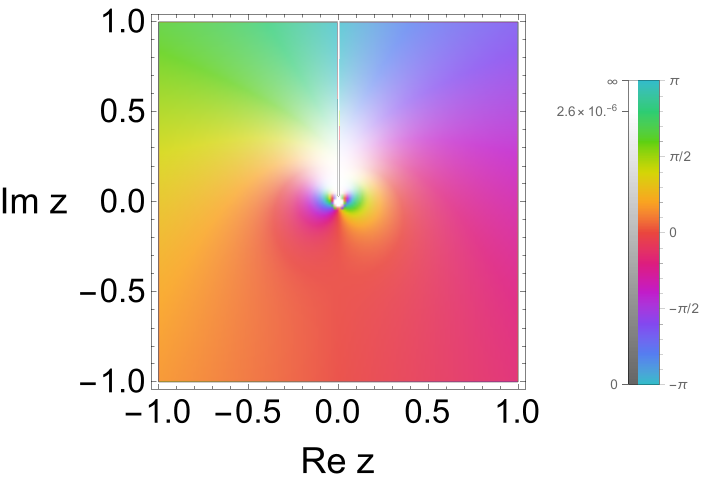}
\caption{$(iz)^{-12}U(17,12,i/z)$}   
\end{subfigure}\hfil
\begin{subfigure}{0.5\textwidth}
    \includegraphics[scale=0.7]{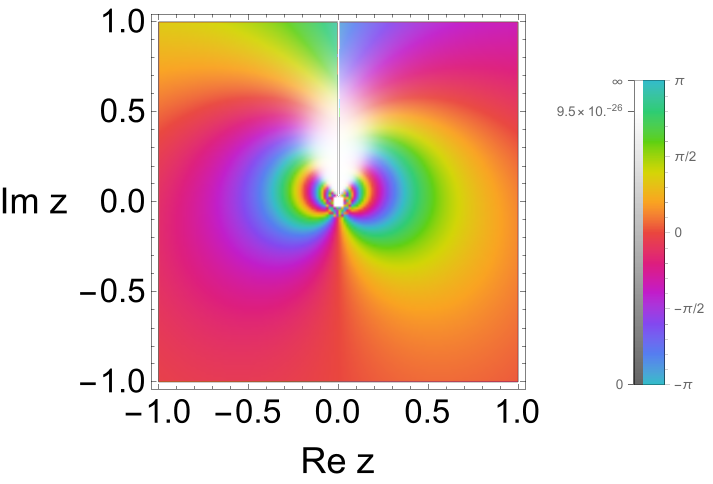}
\caption{$(iz)^{-12}U(32,12,i/z)$}   
\end{subfigure}\hfil
\begin{subfigure}{0.5\textwidth}
    \includegraphics[scale=0.7]{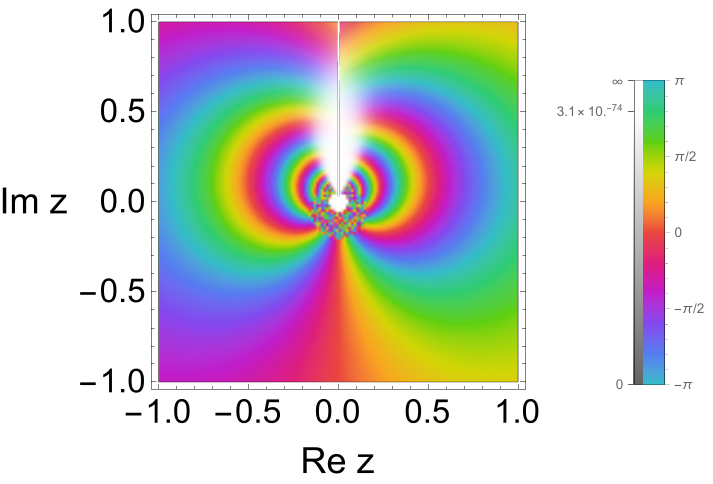}
\caption{$(iz)^{-12}U(62,12,i/z)$}   
\end{subfigure}\hfil
\caption{Plot in the complex plane of $(iz)^{-12}U(k+12,12,i/z)$ ($k=1,5,20,50$), in terms of which the response function $\eta_{2}( \widehat{\Omega})$ is expressed (see Eq.~\eqref{eq:eta-trot-2}).
The colorbar indicates the phase of the Tricomi function, while the gray-scale bar (interpreted as opacity, white being clear and black being opaque) indicates its magnitude.}
\label{fig:U12}
\end{figure}

Collecting the above results, we can compute the shear response function as
\begin{equation}
\begin{aligned}
\widetilde{\pi}^{\mu \nu}_{2} &= 
\frac{2 \beta}{15} \widetilde{\sigma}^{\mu \nu} \sum_{k_{1}=0}^{\infty} \sum_{k_{2}=0}^{\infty} \sum_{j = 0}^{k_{1} + k_{2}} \int_{0}^{\infty} dx  a_{j}^{(k_{1}, k_{2})} b_{k_{1}}(x, \Omega) b_{k_{2}}(x, \Omega)
\exp\left( \frac{x}{2} \chi_{02} + \frac{x}{2} \chi_{k_{1},2} \right)  
\int dP \left( \Delta^{\alpha \beta} p_{\alpha} p_{\beta} \right)^{2} f_{0 {\bf p}} 
 L^{(5)}_{j, {\bf p}}  L^{(5)}_{0, {\bf p}} 
\\
&
=
\frac{16}{g \beta^{3}}\widetilde{\sigma}^{\mu \nu} \sum_{k=0}^{\infty} \binom{k + 5}{k} \int_{0}^{\infty} d\widehat{x}  \frac{ [i (\widehat{\Omega}/2) \widehat{x}]^{2 k} }{\left(1 + i (\widehat{\Omega}/2) \widehat{x} \right)^{2k + 12}} 
\exp\left( \frac{x}{2} \chi_{02} + \frac{x}{2} \chi_{k,2} \right)  
\equiv
2 \eta_{2}(\widehat{\Omega}) \widetilde{\sigma}^{\mu \nu}, 
\end{aligned}    
\end{equation}
where we defined $\widehat{x} = x \mathcal{M}$ and we can identify 
\begin{equation}
\label{eq:eta-trot-2}
\begin{aligned}
\eta_{2}(\widehat{\Omega})
&=
\frac{16}{g \beta^{3}}\sum_{k=0}^{\infty}    \binom{k + 5}{k}
\Gamma(2 k + 1) \frac{1}{i \widehat{\Omega}}  U\left(2 k + 1, -10,  \frac{|\widehat{\chi}_{02}| + |\widehat{\chi}_{k_{1},2}|}{i \widehat{\Omega}}  \right)
\\
&
=
\frac{16}{g \beta^{3}}\sum_{k=0}^{\infty} \binom{k + 5}{k}
\frac{\Gamma(2 k + 1)}{i \widehat{\Omega}} \left(\frac{|\widehat{\chi}_{02}| + |\widehat{\chi}_{k,2}|}{i \widehat{\Omega}} \right)^{11}
U\left(2 k + 12, 12,  \frac{|\widehat{\chi}_{02}| + |\widehat{\chi}_{k,2}|}{i \widehat{\Omega}}  \right),
\end{aligned}    
\end{equation}
and have employed the property in Eq.~\eqref{eq:U-eq-zn-U}. In contrast to the first-order Trotterization truncation, the response function is now an infinite sum of Tricomi confluent hypergeometric functions. Nevertheless, the cut of each term is at the same semi-axis, as evidenced by the plots in Figs.~\ref{fig:U12}, where the typical behavior of the functions $(iz)^{-12}U(2k+12,12,i/z)$ appearing in Eq.~\eqref{eq:eta-trot-2} are displayed in the complex plane. In these figures, it is seen that as one increases the value of $k$, the behavior of the phase of the complex function becomes more complicated and the absolute value of the function becomes significantly smaller (as indicated by the black and white color gradient in the left-hand side of the color legend). Indeed, for $k=1$ the absolute value of the function is below $0.045$, whereas for $k=50$, it cannot exceed values larger than $10^{-74}$.


\subsubsection*{Hydrodynamic series}

In the limit where $\Omega \to 0$, in a manner analogous to the previous subsection, we derive that the asymptotic hydrodynamic series is
\begin{equation}
\begin{aligned}
&
\eta_{2}(\widehat{\Omega})
\sim
\frac{16}{g \beta^{3}}\sum_{k,s=0}^{\infty} \binom{k + 5}{k}
\frac{\Gamma(2 k + 1)}{i \widehat{\Omega}} 
\frac{{\left(2k+12\right)_{s}}{%
\left(2k+1\right)_{s}}}{s!}
\left(-\frac{i \widehat{\Omega}}{|\widehat{\chi}_{02}| + |\widehat{\chi}_{k,2}|} \right)^{2k+s+1}
,
\end{aligned}    
\end{equation}
where $\eta_{2}(0) \sim 48/(g \beta^{3})$, which also agrees with the expressions  computed in Refs.~\cite{Denicol:2022bsq, Rocha:2023hts}, and coincides with the first Trotterization truncation order expression. Up to subleading order, the contribution to the response function is
$\eta_{2}(\widehat{\Omega})
\sim
48/(g \beta^{3})
\left[
1 -  36 i \widehat\Omega\right]
$, which also coincides with the leading Trotterization expansion. Furthermore, the branch cut discontinuity can be computed, and it also possesses an asymptotic series expression that contains an essential singularity near $\widehat{\Omega} = 0$.

\subsection{Third and higher Trotterization truncations}

For the sake of completeness, we now discuss the higher-order Trotterization truncations. Then, we shall compute Eq.~\eqref{eq:pi-n} following steps similar to the ones discussed in the previous subsections,   
\begin{equation}
\begin{aligned}
&
\left[ \exp\left(- i \Omega \frac{x}{n} E_{\bf p} \right)  \exp\left( \frac{x}{n} \Hat{L} \right) \right]^{n}
p^{\langle \alpha} p^{\beta \rangle}
\\
&
=
\sum_{k_{1}=0}^{\infty} \sum_{k_{2}=0}^{\infty} \sum_{k_{3} = 0}^{k_{1} + k_{2}} a_{k_{3}}^{(k_{1}, k_{2})} b_{k_{1}}(x, \Omega) b_{k_{2}}(x, \Omega)
\exp\left( \frac{x}{n} \chi_{02} + \frac{x}{n} \chi_{k_{1},2} \right)
\left[ \exp\left(- i \Omega \frac{x}{n} E_{\bf p} \right)  \exp\left( \frac{x}{n} \Hat{L} \right) \right]^{n-2}
 L^{(5)}_{k_{3}, {\bf p}} p^{\langle \alpha} p^{\beta \rangle}
\\
&
=
\sum_{k_{1}=0}^{\infty} \sum_{k_{2}=0}^{\infty} \sum_{k_{3} = 0}^{k_{1} + k_{2}}
\sum_{k_{4}=0}^{\infty} \sum_{k_{5} = 0}^{k_{3} + k_{4}}
a_{k_{5}}^{(k_{3}, k_{4})} b_{k_{4}}(x, \Omega) a_{k_{3}}^{(k_{1}, k_{2})} b_{k_{1}}(x, \Omega) b_{k_{2}}(x, \Omega)
\exp\left( \frac{x}{n} \chi_{02} + \frac{x}{n} \chi_{k_{1},2} + \frac{x}{n} \chi_{k_{3},2} \right)
\\
&
\times
\left[ \exp\left(- i \Omega \frac{x}{n} E_{\bf p} \right)  \exp\left( \frac{x}{n} \Hat{L} \right) \right]^{n-3}
 L^{(5)}_{k_{5}, {\bf p}} p^{\langle \alpha} p^{\beta \rangle}
\\
&
\vdots
\\
&
=
\sum_{k_{1}=0}^{\infty}
\sum_{k_{2}=0}^{\infty}
\sum_{k_{3}=0}^{k_{1}+k_{2}}
\sum_{k_{4}=0}^{\infty}
\sum_{k_{5}=0}^{k_{3}+k_{4}}
 \cdots  
\sum_{k_{2n-2}=0}^{\infty} 
\sum_{k_{2n-1}=0}^{k_{2n-3}+k_{2n-2}} 
a_{k_{2n-1}}^{(k_{2n-3},k_{2n-2})} b_{k_{2n-2}}(x, \Omega) \cdots 
a^{(k_{3},k_{4})}_{k_{5}} b_{k_{4}}(x, \Omega)
a^{(k_{1},k_{2})}_{k_{3}} b_{k_{2}}(x, \Omega) b_{k_{1}}(x, \Omega) \\
&
\times 
\exp\left[ \frac{x}{n} \left(\chi_{02} + \sum_{j=1}^{n-1} \chi_{k_{2j-1} 2} \right) \right] L_{k_{2n -1}, {\bf p}}^{(5)} p^{\langle \alpha} p^{\beta \rangle}.
\end{aligned}    
\end{equation}
In the first step above, we have performed computations analogous to Eq.~\eqref{eq:second-ord-trot-trun} in the last subsection. In the second step, we see that the action of each $\exp\left(- i \Omega \frac{x}{n} E_{\bf p} \right)  \exp\left( \frac{x}{n} \Hat{L} \right)$ operator leads to the emergence of an infinite sum in $k_{4}$ and a finite sum in $k_{5}$. Repeating this procedure until there are no $\exp\left(- i \Omega \frac{x}{n} E_{\bf p} \right)  \exp\left( \frac{x}{n} \Hat{L} \right)$ operators left, we have the final result. We remark that, in the $n$-th Trotterization truncation, analogous to Eq.~\eqref{eq:eiE-b}, we have
\begin{equation}
\begin{aligned}
&
b_{k}(x, \Omega) = b_{k}\left(\frac{\Omega x}{n \beta}\right) = \frac{ (i \frac{\Omega x}{n \beta})^{k} }{\left(1 + i \frac{\Omega x}{n \beta} \right)^{6+k}}.
\end{aligned}    
\end{equation}

Similarly to the previous sections, we derive the shear-stress response function as 
\begin{equation}
\begin{aligned}
&
\widetilde{\pi}^{\mu \nu}_{n}
=
2 \eta_{n}(\widehat{\Omega})
\widetilde{\sigma}^{\mu \nu},
\end{aligned}    
\end{equation}
where we identify
{\allowdisplaybreaks
\begin{align}
&
\notag
\eta_{n}(\widehat{\Omega})
=
\frac{8}{g\beta^{3}} \widetilde{\sigma}^{\mu \nu} \sum_{k_{1}=0}^{\infty}
\sum_{k_{2}=0}^{\infty}
\sum_{k_{3}=0}^{k_{1}+k_{2}}
\sum_{k_{4}=0}^{\infty}
\sum_{k_{5}=0}^{k_{3}+k_{4}}
 \cdots  
 \sum_{k_{2n-4}=0}^{\infty}
\sum_{k_{2n-3}=0}^{k_{2n-5}+k_{2n-4}} 
\sum_{k_{2n-2}=0}^{\infty}  a_{0}^{(k_{2n-3},k_{2n-2})} a_{k_{2n-3}}^{(k_{2n-5},k_{2n-4})} \cdots 
a^{(k_{3},k_{4})}_{k_{5}} 
a^{(k_{1},k_{2})}_{k_{3}}
\\
&
\notag
\times
\int_{0}^{\infty} d\widehat{x} \frac{ [i (\widehat{\Omega}/n) \widehat{x}]^{k_{1} + \sum_{j=1}^{n-1} k_{2j}} }{\left(1 + i (\widehat{\Omega}/n) \widehat{x} \right)^{6 n + k_{1} + \sum_{j=1}^{n-1} k_{2j}}}   \exp\left[ \frac{\widehat{x}}{n} \left(\widehat{\chi}_{02} + \sum_{j=1}^{n-1} \widehat{\chi}_{k_{2j-1} 2} \right) \right]
\\
&
\notag
=
\frac{8}{g \beta^{3}}\lim_{n \to \infty} 
\sum_{k_{1}=0}^{\infty}
\sum_{k_{2}=0}^{\infty}
\sum_{k_{3}=0}^{k_{1}+k_{2}}
\sum_{k_{4}=0}^{\infty}
\sum_{k_{5}=0}^{k_{3}+k_{4}}
 \cdots  
 \sum_{k_{2n-4}=0}^{\infty}
\sum_{k_{2n-3}=0}^{k_{2n-5}+k_{2n-4}}
\binom{k_{2n-3} + 5}{k_{2n-3}} a_{k_{2n-3}}^{(k_{2n-5},k_{2n-4})} \cdots 
a^{(k_{3},k_{4})}_{k_{5}} 
a^{(k_{1},k_{2})}_{k_{3}}
\\
&
\notag
\times
\frac{n\Gamma(K_{n}+1)}{i \widehat{\Omega}} U\left(K_{n}+1,2-6n, \frac{\vert \widehat{\chi}_{02} \vert + \sum_{j=1}^{n-1} \vert \widehat{\chi}_{k_{2j-1} 2} \vert}{i \widehat{\Omega}} \right)
\\
&
=
\notag
\frac{8 n}{g \beta^{3}} 
\sum_{k_{1}=0}^{\infty}
\sum_{k_{2}=0}^{\infty}
\sum_{k_{3}=0}^{k_{1}+k_{2}}
\sum_{k_{4}=0}^{\infty}
\sum_{k_{5}=0}^{k_{3}+k_{4}}
 \cdots  
\sum_{k_{2n-4}=0}^{\infty}
\sum_{k_{2n-3}=0}^{k_{2n-5}+k_{2n-4}}
\binom{k_{2n-3} + 5}{k_{2n-3}} a_{k_{2n-3}}^{(k_{2n-5},k_{2n-4})} \cdots 
a^{(k_{3},k_{4})}_{k_{5}} 
a^{(k_{1},k_{2})}_{k_{3}}
\\
&
\times
\left(
\frac{\vert \widehat{\chi}_{02} \vert + \sum_{j=1}^{n-1} \vert \widehat{\chi}_{k_{2j-1} 2} \vert}{i \widehat{\Omega}} \right)^{6n-1}
\frac{\Gamma(K_{n}+1)}{i \widehat{\Omega}}  U\left(K_{n}+6n,6n, \frac{\vert \widehat{\chi}_{02} \vert + \sum_{j=1}^{n-1} \vert \widehat{\chi}_{k_{2j-1} 2} \vert}{i \widehat{\Omega}} \right),
\end{align}    
where $K_{n} = (k_{1} + \sum_{j=1}^{n-1} k_{2j}) \delta_{k_{2n-3}, k_{2n-2}} = k_{1} + \sum_{j=1}^{n-2} k_{2j} + k_{2n-3} $. Thus, since all of the terms possess a branch cut discontinuity, it is expected that the response function also does.

In the present case, the hydrodynamic series expansion leads to the following asymptotic series expression for the response function,
\begin{equation}
\begin{aligned}
&
\eta_{n}(\widehat{\Omega})
\sim
\frac{8 n}{g \beta^{3}} 
\sum_{k_{1}=0}^{\infty}
\sum_{k_{2}=0}^{\infty}
\sum_{k_{3}=0}^{k_{1}+k_{2}}
\sum_{k_{4}=0}^{\infty}
\sum_{k_{5}=0}^{k_{3}+k_{4}}
 \cdots  
\sum_{k_{2n-4}=0}^{\infty}
\sum_{k_{2n-3}=0}^{k_{2n-5}+k_{2n-4}}
\sum_{s=0}^{\infty}
\binom{k_{2n-3} + 5}{k_{2n-3}} a_{k_{2n-3}}^{(k_{2n-5},k_{2n-4})} \cdots 
a^{(k_{3},k_{4})}_{k_{5}} 
a^{(k_{1},k_{2})}_{k_{3}}
\\
&
\times
\frac{\Gamma(K_{n}+1)}{i \widehat{\Omega}} 
\frac{{\left(K_{n}+6n\right)_{s}}{%
\left(K_{n}+1\right)_{s}}}{s!}
\left(
-\frac{i \widehat{\Omega}}{\vert \widehat{\chi}_{02} \vert + \sum_{j=1}^{n-1} \vert \widehat{\chi}_{k_{2j-1} 2} \vert} \right)^{K_n+s+1},
\end{aligned} 
\end{equation}
where also at next to leading order in the asymptotic $\widehat\Omega \to 0$, $\eta_{n}(\widehat\Omega) \sim 48/(g \beta^{3}) \left[
1 -  36 i \widehat\Omega\right]$ (details in Appendix \ref{apn:a-b-coeffs}). Thus, at any given order in the Trotterization truncation, the zeroth and first-order contributions to the asymptotic series coincide, respectively, with the expression for shear viscosity and shear relaxation time computed in Refs.~\cite{Rocha:2023hts}. Moreover, one also sees that the general qualitative behavior of the response function can be captured by the first and the second Trotterization truncations, with higher-order corrections providing minor quantitative improvements.

\section{DISCUSSION}

In this paper, we showed analytically that the shear-response function for the $\lambda \varphi^{4}$ theory possesses a branch cut for $ i 0^{+} < \Omega < i \infty$. This is a symptom of (infinitely many) long-lived non-hydrodynamic modes in the theory. These modes arise from the total cross-section for this system, which has the functional form $\sigma(s) = g/s$, which becomes negligible for particles with large energy in the center of the momentum frame. The branch-cut extending to the origin provides an explicit example where the method proposed in Ref.~\cite{Denicol:2021} for determining the relaxation time cannot be applied. 
We emphasize that this does not imply the break down of hydrodynamics or kinetic theory. 

In Sec.~\ref{sec:analyt-trotter}, we showed that for all of the Trotterization truncations, the asymptotic expression for the response function in $\widehat\Omega$, is such that $\pi^{\mu \nu} = 2 \eta_{n}(\widehat\Omega) \sigma^{\mu \nu}$, with  $\eta_{n}(\widehat\Omega) \sim 48/(g \beta^{3}) \left[
1 -  36 i \widehat\Omega\right]$. This suggests that the equation of motion for the shear-stress tensor may be approximated by an relaxation equation in time, $[\tau_{\pi} D + 1 ]\pi^{\mu \nu} \simeq 2 \eta \sigma^{\mu \nu}$, with $\tau_{\pi} = 72/(g n_{0} \beta^{2})$ and $\eta = 48/(g \beta^{3})$, which coincides with the transport coefficients computed in Ref.~\cite{Rocha:2023hts} from the Boltzmann equation using the order of magnitude approach \cite{struchtrup2004stable,Fotakis:2022usk}. Finally, since the long-lived non-hydrodynamic modes originate from the very weakly interacting high-energy colliding particles, it can be argued that the system is partitioned into two sets: the first set contains these long-lived high-energy non-hydrodynamic modes and a second set containing the low-energy underlying hydrodynamic degrees of freedom \cite{Gavassino:2024rck}. Our results indicate that these high-energy modes, even though long-lived, do not contribute to the transport properties of the system due to their weak interaction.

This paper discusses massless $\lambda \varphi^4$ theory in the kinetic regime, i.e. small coupling, with classical statistics and tree-level cross-section. Effects due to quantum fluctuations and/or strong coupling are beyond the scope of this paper. We note that strong coupling results for the non-hydrodynamic modes can be obtained using holography \cite{Kovtun:2005ev}, where the spectrum was shown to be gapped with poles with nonzero real and imaginary parts. We note that radiative corrections to the collision amplitudes were not considered in this paper and can i,n principle alter the dynamics of the system affecting the existence of the branch cut singularity. Besides that, this singularity may also not be present in systems with large occupancy numbers. These effects, which are relevant when quantum corrections are appreciable, will be investigated in future works.

\vspace{0.5cm}

\section{Conclusion}
\label{sec:Conclusion}

In the present work, we have analyzed the linear response function of the shear-stress tensor, $\pi^{\mu \nu}$, within kinetic theory for a system of classical, scalar ultrarelativistic particles with quartic self-interactions. Within this model, the eigenfunctions and eigenvalues of the linearized collision term, $\hat{L}$, are known in analytical form. This enabled us to show in two independent ways that the response function for the shear stress, which has the form $\widetilde\pi^{\mu \nu} = 2 \eta(\widehat{\Omega}) \widetilde\sigma^{\mu \nu}$, possesses a branch-cut singularity in the Fourier variable $\Omega = q_{\mu} u^{\mu}$ for homogeneous perturbations around local equilibrium ($\widehat{\Omega} = 2\Omega/(g n_{0}\beta^{2})$). The first method we used involved obtaining an infinite system of linear equations for the moments of the $f_{\bf p}$ distribution with respect to the eigenfunctions of the linearized Boltzmann equation in Fourier space, Eq.~\eqref{eq:main-linear-sys}. One such moment is the shear-stress tensor. Then, truncating this infinite tower of equations, one can obtain an analytical expression for the $\pi^{\mu \nu}$-truncated response function since the matrix defining it is tridiagonal. Assessing the behavior of the singularities of the system at increasing finite orders (up to $N=750$), we have shown that the poles lie in the  $\Im \widehat\Omega > 0$, $\Re \widehat\Omega = 0$ semi-axis (due to our choices of Fourier and metric signature). The pole closest to the origin approaching $\widehat\Omega = 0$ is such  that $\min \widehat{\Omega} \propto 1/N^{0.996}$. Furthermore, we see that the poles get closer to one another since the average relative distance between the 15 poles closest to the origin, $\langle \Delta \Omega \rangle_{15}$, behaves as $\langle \Delta \Omega \rangle_{15} \propto 1/N^{1.794}$.    

Our second approach involved Trotterization methods to formally invert the linear inhomogeneous integral equation, Eq.~\eqref{eq:main-linear-sys}, in the shear sector. Then, the inverse of the operator is expressed as a limit in a discrete variable of products of exponentials involving the momentum-space function operators $\hat{L}$ and $E_{\bf p}$ and an integral in an auxiliary variable. Each iteration in the discrete variable defines a Trotterization truncation. Since the properties of the action of both operators on the eigenbasis functions $L_{n}^{(5)} p^{\langle \alpha} p^{ \beta \rangle}$ are well-known, and the integrals in momentum space and the auxiliary variable can be performed, the shear response function can be computed. Then, one finds that this function of $\widehat\Omega$ can be expressed in terms of a series involving Tricomi confluent hypergeometric functions, which indeed possess a branch cut singularity along the $\Im \widehat\Omega > 0$, $\Re \widehat\Omega = 0$ semi-axis for all of the terms of series.

Next, we showed that for all of the Trotterization truncations, the asymptotic expression for the response function recovers the hydrodynamic transport coefficients in the $\Omega \to 0$ limit. This happens even though $\widehat\Omega = 0$ is a branch point of the response function. This result highlights a subtle aspect of the derivation of \textit{relativistic} dissipative theories of hydrodynamics: the derivations of Refs.~\cite{Denicol:2022bsq,Rocha:2023hts} take into account only properties related to $\Hat{L}$ itself. In contrast, the full linear-response dynamics derived here emerge from the non-commutativity between $\Hat{L}$ and $E_{\bf p}$, from which the branch cut structure emerges. In Ref.~\cite{Denicol:2012cn}, however, the relaxation times of the dissipative currents coincide with the longest-lived non-hydrodynamic mode. This is not the case in the present paper since this timescale is usually interpreted as the pole of the linear response function.

The results demonstrated in this paper also imply that previous analyses \cite{Moore:2018mma,Ochsenfeld:2023wxz} showing this behavior for the non-hydrodynamic modes are not due to quantum statistics or quantum effects. Rather, our results show that this property emerges, at least in a qualitative manner, from the underlying non-equilibrium dynamics for the given interaction even at the classical level. The origin of such a feature is that the cross-section for the $\varphi^{4}$ theory is negligible for large (center-of-momentum) energies. Then, large energy modes overpopulate evolution at late times, thus giving rise to infinitely many long-lived modes. Since the value of $\Im \widehat\Omega$ for a singularity is the reciprocal of the relaxation time of the mode, the above-mentioned overpopulation manifests as a branch cut that touches the origin. In the future, it would be important to investigate these properties in more general non-equilibrium perturbation configurations and to consider other sectors beyond that of shear. Then, it would be possible to assess how the higher-order transport coefficients capture the properties of the linear response function. Additionally, it would be interesting to understand if branch cuts could also be present in the shear stress correlators of systems with other types of interactions. Finally, since quantum statistical effects are not crucial for the emergence of these branch cuts, it is reasonable to assume that their presence stems mainly from the properties of the particle's interactions. If that is the case, it should be possible to find general conditions concerning the properties of the interactions that imply the presence of branch cuts. 

NOTE ADDED: While this paper was being finished, we became aware of a general theorem by L.~Gavassino presented in \cite{Gavassino:2024rck} that can be used to determine when a given kinetic system will display the type of branch cut discussed here. Indeed, Gavassino showed that $\lambda \varphi^{4}$ theory satisfies the requirements of the theorem, which provides solid proof for the existence of branch cuts in such a system.

\section*{ACKNOWLEDGEMENTS}

The authors thank L.~Gavassino for fruitful discussions and for sharing with us the results of his upcoming work.  G.~S.~R.~is supported by Vanderbilt University and in part by the U.S. Department of Energy, Office of Science under Award Number DE-SC-002434. J.~N.~was partially supported by the U.S. Department of Energy, Office of Science, Office for Nuclear Physics under Award No. DE-SC0023861. I.~D.~was supported by the Deutsche Forschungsgemeinschaft (DFG, German Research Foundation) through the CRC-TR 211 `Strong-interaction matter under extreme conditions'– project number 315477589 -- TRR 211. G.~S.~D.~acknowledges CNPq as well as Fundação Carlos Chagas Filho de Amparo à Pesquisa do Estado do Rio de Janeiro (FAPERJ), grant No.~E-26/202.747/2018. K.I. is supported by the National Science Foundation under Grant No. NSF PHYS-2316630. Any opinions, findings, and conclusions or recommendations expressed in this material are those of the author(s) and do not necessarily reflect the views of the National Science
Foundation.

\appendix

\section{The coefficients $a_{j}^{(k_{1}, k_{2})}$, $b_{k}(x, \Omega)$ and $n$-th Trotterization truncation}
\label{apn:a-b-coeffs}

In this Appendix, we shall provide details to obtain the coefficients $a_{j}^{(k_{1}, k_{2})}$ and $b_{k}(x, \Omega)$ employed, respectively in Eqs.~\eqref{eq:LL-a-coeffs} and \eqref{eq:eiE-b}. We start with $a_{j}^{(k_{1}, k_{2})}$ which correspond to the coefficients in the expansion of the product of two Laguerre polynomials $L_{k_{1},{\bf p}}^{(5)} L_{k_{2},{\bf p}}^{(5)}$ as a linear combination Laguerre polynomials,
\begin{equation}
\begin{aligned}
&
 L_{k_{1}, {\bf p}}^{(5)} L_{k_{2}, {\bf p}}^{(5)}
 =
\sum_{j_{1}=0}^{k_{1} + k_{2}} a^{(k_{1},k_{2})}_{j_{1}} L_{j_{1}, {\bf p}}^{(5)}.
\end{aligned}   
\end{equation}
Integrating both sides of the above equation with $(\Delta_{\mu \nu} p^{\mu} p^{\nu})^{2} L_{j_{1}, {\bf p}}^{(5)} f_{0{\bf p}}$, we have 
\begin{equation}
\begin{aligned}
&
A_{j_{1}}^{(2)}
a^{(k_{1},k_{2})}_{j_{1}}
=
\frac{n_{0}}{2 \beta^{3}}
\int_{0}^{\infty} du \ u^{5} L_{j_{1}, {\bf p}}^{(5)} L_{k_{1}, {\bf p}}^{(5)} L_{k_{2}, {\bf p}}^{(5)}
=
\frac{n_{0}}{2 \beta^{3}}
\sum_{s = 0}^{j_{1}}
\frac{(-1)^{s}}{s!} \binom{j_{1} + 5}{j_{1} - s}
\int_{0}^{\infty} du \ u^{5+s} L_{k_{1}, {\bf p}}^{(5)} L_{k_{2}, {\bf p}}^{(5)},
\end{aligned}    
\end{equation}
where in the last step, we have used the fact that the polynomial coefficients for Laguerre coefficients are known
\begin{equation}
\label{eq:L-polyn-coeffs}
L_{j_{1}}^{(\alpha)}(x) =   \sum_{s=0}^{j_{1}} \frac{(-1)^{s}}{s!} \binom{j_{1} + \alpha}{j_{1}-s} 
 x^{s}.
\end{equation}
For $j_{1} = 0$, using equation \eqref{eq:orth-laguerre} one can readily derive that 
\begin{equation}
\begin{aligned}
&
a_{0}^{(k_{1}, k_{2})} 
=
\binom{k_{1} + 5}{k_{1}} \delta_{k_{1}, k_{2}}.
\end{aligned}    
\end{equation}
For $j_{1} > 0$ one may repeatedly employ the identity (see also Eq.~\eqref{eq:lag-prop-E}) $u L_{n, {\bf p}}^{(5)} 
=
-
(n+1) L_{n+1, {\bf p}}^{(5)}
+
2(n+3) L_{n, {\bf p}}^{(5)}
-
(n+5) L_{n-1, {\bf p}}^{(5)}$ to derive  an expression of $u^{s} L_{k_{2}, {\bf p}}^{(5)}$ as a linear combination of the Laguerre polynomials $L_{k_{2}-s, {\bf p}}^{(5)}$ (for $k_{2} \geq s$, otherwise from $L_{0, {\bf p}}^{(5)}$) until $L_{k_{2}+s, {\bf p}}^{(5)}$, which then implies that $a_{j_{1}}^{(k_{1}, k_{2})} $ is a linear combination of $\delta_{k_{1}, k_{2}-j_{1}}$, $\delta_{k_{1}, k_{2}-j_{1}+1}$, $\cdots$, $\delta_{k_{1}, k_{2}}$, $\cdots$,  $\delta_{k_{1}, k_{2}+j_{1}}$. Alternatively, a more compact and more convenient form for the $a_{j_{1}}^{(k_{1}, k_{2})}$ coefficients can be derived 
\begin{equation}
\begin{aligned}
&
L_{k_{1}, {\bf p}}^{(5)} L_{k_{2}, {\bf p}}^{(5)}
=
\sum_{s=0}^{k_{1}}
\sum_{t=0}^{k_{2}}
\frac{(-u)^{s+t}}{s!t!} \binom{k_{1} + 5}{k_{1} - s} \binom{k_{2} + 5}{k_{2} - t}
\\
&
=
\sum_{s=0}^{k_{1}}
\sum_{t=0}^{k_{2}}
\sum_{j=0}^{s+t} \frac{(-1)^{s+t+j}}{s!t!} \binom{k_{1} + 5}{k_{1} - s} \binom{k_{2} + 5}{k_{2} - t}
\left(
\begin{array}{c}
 s+t+5 \\
 s+t-j     
\end{array}
\right) L_{j,{\bf p}}^{(5)},
\end{aligned}    
\end{equation}
where, in the first step of the above expression, Eq.~\eqref{eq:L-polyn-coeffs} has been employed, and then the relation \cite{NIST:DLMF,gradshteyn2014table}
\begin{equation}
\begin{aligned}
&
u^{k} = k! \sum_{j=0}^{k}(-1)^{j} \binom{k + 5}{k - j} L^{(5)}_{j}(u).
\end{aligned}    
\end{equation}
From that, manipulations allow us to identify
\begin{equation}
\begin{aligned}
&
a_{j}^{(k_{1}, k_{2})}
=
(k_{1}+5)!(k_{2}+5)!\sum_{s=0}^{k_{1}}
\sum_{t=j}^{k_{1} + k_{2}} \frac{(-1)^{t+j_{1}}}{(k_{1} + k_{2} - t)!(t+10)!}
\left(
\begin{array}{c}
 k_{1} + k_{2} - t \\
 k_{1} - s   
\end{array}
\right)
\left(
\begin{array}{c}
 t + 10\\
 s + 5    
\end{array}
\right)
\left(
\begin{array}{c}
 t \\
 s     
\end{array}
\right)
\left(
\begin{array}{c}
 t+5 \\
 t-j    
\end{array}
\right).
\end{aligned}    
\end{equation}
Thus, since given two integers $a$ and $b$ the term $\binom{a}{b}$ gives zero whenever $b<0$ or whenever $a-b < 0$, the fact that $a_{j_{1}}^{(k_{1}, k_{2})} $ is a linear combination of $\delta_{k_{1}, k_{2}-j_{1}}$, $\delta_{k_{1}, k_{2}-j_{1}+1}$, $\cdots$, $\delta_{k_{1}, k_{2}}$, $\cdots$,  $\delta_{k_{1}, k_{2}+j_{1}}$ can be recovered.

On the other hand, the coefficients $b_{nk}(x, \Omega)$ express phases as the linear combination of Laguerre polynomials   
\begin{equation}
\begin{aligned}
&
 \exp\left(- i \Omega \frac{x}{n} E_{\bf p} \right)  = \sum_{k=0}^{\infty} b_{k}(x,\Omega) L_{k, {\bf p}}^{(5)}.
\end{aligned}    
\end{equation}
Once again, integrating both sides of the above equation with $(\Delta_{\mu \nu} p^{\mu} p^{\nu})^{2} L_{k{\bf p}}^{(5)} f_{0{\bf p}}$, we have
\begin{equation}
\begin{aligned}
b_{k}(x,\Omega) 
&=
\frac{n_{0}}{2\beta^{3}A_{k}^{(2)}}\int du \ u^{5} \exp\left[- \left(1 + i \Omega \frac{x}{n \beta} \right) u \right] L_{k}^{(5)}(u)
\\
&
=
\frac{n_{0}}{2\beta^{3}A_{k}^{(2)}} \frac{1}{\left(1 + i \Omega \frac{x}{n \beta} \right)^{6}}\int dw \ w^{5} \exp\left[- w \right] L_{k}^{(5)} \left( \frac{w}{1 + i \Omega \frac{x}{n \beta}} \right)
\\
&
=
\frac{n_{0}}{2\beta^{3}A_{k}^{(2)}} \frac{1}{\left(1 + i \Omega \frac{x}{n \beta} \right)^{6}} \sum_{j=0}^{k} \frac{(-1)^{j}}{j!} \binom{k + 5}{k-j} \left( \frac{1}{1 + i \Omega \frac{x}{n \beta}} \right)^{j}\int dw \ w^{5+j} \exp\left[- w \right] 
\\
&
=
\frac{ (i\frac{\Omega }{n \beta} x)^{k} }{\left(1 + i \frac{\Omega }{n \beta} x \right)^{6+k}}. 
\end{aligned}    
\end{equation}
In the first to the second equality above, we made the change of variables $u \mapsto w = \left(1 + i \Omega \frac{x}{n \beta} \right) u$. From the second to the third expression, we employed identity \eqref{eq:L-polyn-coeffs}. In the final step, we employed the summation identity $\sum_{j=0}^k \frac{(-1)^j}{\lambda ^j (j! (k-j)!)} = (1/k!)(1 - 1/\lambda)^{k}$ \cite{gradshteyn2014table}.

\subsection*{Details of the hydrodynamic expansion in the $n$-th Trotterization truncation}

In Sec.~\ref{sec:analyt-trotter}, we have shown that the $n$-th Trotterization truncation of the shear-stress response function can be expressed as
\begin{equation}
\begin{aligned}
&
\widetilde{\pi}^{\mu \nu}_{n}
=
2 \eta_{n}(\widehat{\Omega})
\widetilde{\sigma}^{\mu \nu},
\end{aligned}    
\end{equation}
where the coefficient $\eta_{n}(\widehat{\Omega})$ admits the following asymptotic expansion, 
\begin{equation}
\label{eq:resp-fun-n-trot-apn}
\begin{aligned}
&
\eta_{n}(\widehat{\Omega})
\sim
\frac{8 n}{g \beta^{3}} 
\sum_{k_{1}=0}^{\infty}
\sum_{k_{2}=0}^{\infty}
\sum_{k_{3}=0}^{k_{1}+k_{2}}
\sum_{k_{4}=0}^{\infty}
\sum_{k_{5}=0}^{k_{3}+k_{4}}
 \cdots  
\sum_{k_{2n-4}=0}^{\infty}
\sum_{k_{2n-3}=0}^{k_{2n-5}+k_{2n-4}}
\sum_{s=0}^{\infty}
\binom{k_{2n-3} + 5}{k_{2n-3}} a_{k_{2n-3}}^{(k_{2n-5},k_{2n-4})} \cdots 
a^{(k_{3},k_{4})}_{k_{5}} 
a^{(k_{1},k_{2})}_{k_{3}}
\\
&
\times
\frac{\Gamma(K_{n}+1)}{i \widehat{\Omega}} 
\frac{{\left(K_{n}+6n\right)_{s}}{%
\left(K_{n}+1\right)_{s}}}{s!}
\left(
-\frac{i \widehat{\Omega}}{\vert \widehat{\chi}_{02} \vert + \sum_{j=1}^{n-1} \vert \widehat{\chi}_{k_{2j-1} 2} \vert} \right)^{K_n+s+1},
\end{aligned} 
\end{equation}
where $K_{n} = k_{1} + \sum_{j=1}^{n-2} k_{2j} + k_{2n-3} $.
The zero-th order contribution for this asymptotic series can be computed by taking $s = 0$ and $K_n = 0$. The latter implies that $k_{1} = 0, k_{2} = 0, k_{4} = 0, \cdots, k_{2n-4} = 0, k_{2n-3} = 0$. In this case, the product of $a$ coefficients simplifies
\begin{equation}
\label{eq:a-deltas-prod}
\begin{aligned}
&
a_{0}^{(k_{2n-5},0)} 
a^{(k_{2n-7},0)}_{k_{2n-5}} \cdots 
a^{(k_{5},0)}_{k_{7}} 
a^{(k_{3},0)}_{k_{5}} 
a^{(0,0)}_{k_{3}}
=
\delta_{0,k_{2n-5}}
\delta_{k_{2n-5},k_{2n-7}}
\cdots
\delta_{k_{7},k_{5}}
\delta_{k_{5},k_{3}}
\delta_{k_{3},0},
\end{aligned}    
\end{equation}
where the properties \eqref{eq:a-deltas} have been employed. Then, the only term in the summations that remains is the one where also $k_{3} = 0, k_{5} = 0, \cdots, k_{2n-5} = 0$. In this case, all the terms in Eq.~\eqref{eq:resp-fun-n-trot-apn} yield one, except for the one in parentheses. The latter yields $\left(1/(n \vert \widehat{\chi}_{02} \vert) \right) = 6/n$. Thus $\eta_{n}(\widehat{\Omega})
\sim
48/(g \beta^{3})
$ at leading order. 

In the next to leading order calculation, the contributions can come, in principle, from the terms in Eq.~\eqref{eq:resp-fun-n-trot-apn} with $s=1$ and $K_{n} = 0$, or $s=0$ and $K_{n} = 1$. The latter shall not contribute to this order. This can be understood by changing one of the $0$'s in the indices in the $a$-coefficients in the left-hand side of the Eq.~\eqref{eq:a-deltas-prod} to a $1$. Taking as examples $k_{2n-3} = 1$, $k_{6} = 1$, $k_{1} = 1$, we have, respectively,
\begin{equation}
\label{eq:a-deltas-prod-2}
\begin{aligned}
a_{0}^{(k_{2n-5},1)} 
a^{(k_{2n-7},0)}_{k_{2n-5}} \cdots 
a^{(k_{5},0)}_{k_{7}} 
a^{(k_{3},0)}_{k_{5}} 
a^{(0,0)}_{k_{3}}
& =
a_{0}^{(k_{2n-5},1)}
\delta_{k_{2n-5},k_{2n-7}}
\cdots
\delta_{k_{7},k_{5}}
\delta_{k_{5},k_{3}}
\delta_{k_{3},0},
\\
a_{0}^{(k_{2n-5},0)} 
a^{(k_{2n-7},0)}_{k_{2n-5}} \cdots 
a^{(k_{7},0)}_{k_{9}}
a^{(k_{5},1)}_{k_{7}} 
a^{(k_{3},0)}_{k_{5}} 
a^{(0,0)}_{k_{3}}
&=
\delta_{0,k_{2n-5}}
\delta_{k_{2n-5},k_{2n-7}}
\cdots
\delta_{k_{9},k_{7}}
a^{(k_{5},1)}_{k_{7}}
\delta_{k_{5},k_{3}}
\delta_{k_{3},0},
\\
a_{0}^{(k_{2n-5},0)} 
a^{(k_{2n-7},0)}_{k_{2n-5}} \cdots 
a^{(k_{5},0)}_{k_{7}} 
a^{(k_{3},0)}_{k_{5}} 
a^{(1,0)}_{k_{3}}
&=
\delta_{0,k_{2n-5}}
\delta_{k_{2n-5},k_{2n-7}}
\cdots
\delta_{k_{7},k_{5}}
\delta_{k_{5},k_{3}}
a^{(1,0)}_{k_{3}},
\end{aligned}    
\end{equation}
thus, the zeros in the deltas ``propagate'' until the term with $a^{(k_{\#},1)}_{k_{\#}}$ and yield $a^{(0,1)}_{0} = 0$. On the other hand, the term in Eq.~\eqref{eq:resp-fun-n-trot-apn} with $s=1$ and $K_{n} = 0$ yield a product of $a$-coefficients that only contributes with a factor of 1 when all the $k$ indices are zero. The remaining factors in the second line of Eq.~\eqref{eq:resp-fun-n-trot-apn} yield $-(6 i \widehat{\Omega})/(n \vert \widehat{\chi}_{02} \vert^{2}) = 216/n$, which implies that at next to leading order in $\widehat\Omega$, $\eta_{n}(\widehat\Omega) \sim 48/(g \beta^{3}) \left[
1 -  36 i \widehat\Omega\right]$.

\bibliographystyle{apsrev4-1}
\bibliography{liography}

\end{document}